\documentclass[useAMS,usenatbib]{mn2e}

\usepackage{aas_macros}

\usepackage{graphicx} \usepackage{amssymb} \usepackage{amsmath}

\usepackage{graphicx} 
\usepackage{amssymb}
\usepackage{amsmath}
\usepackage{color}

\usepackage{natbib}

\def\msun{\mbox{$M_\odot$}}
\def\ms{\mbox{$M_{\ast}$}}
\def\mg{\mbox{$M_{\rm gas}$}}
\def\mh{\mbox{$M_{h}$}}

\def\lcdm{\mbox{$\Lambda$CDM}}

\def\bt{\mbox{$B/T$}}
\def\zmorph{\mbox{$z_{\rm morph}$}}
\def\zass{\mbox{$z_{\rm assem}$}}

\defcitealias{Zavala+2012}{Paper I}

\title[The growth of galactic bulges through mergers in $\Lambda$CDM haloes revisited II]
{The growth of galactic bulges through mergers in $\Lambda$CDM haloes revisited. II. Morphological mix evolution} 
\author[Avila-Reese et al.] {\parbox{17.5cm}{Vladimir Avila-Reese$^{1}$\thanks{Email: avila@astro.unam.mx}, Jes\'us
          Zavala$^{2,3,4}$, and Ivan Lacerna$^{1}$ 
                  }\vspace{0.3cm}\\ 
        $^1$Instituto de Astronom{\' i}a, Universidad Nacional Aut{\' o}noma
        de M{\'e}xico, A.P. 70-264, 04510, M{\'e}xico, D.F., M{\'e}xico\\
        $^2$Perimeter Institute for Theoretical Physics, 31 Caroline St. N., Waterloo, ON, N2L 2Y5, Canada\\
        $^3$Department of Physics and Astronomy, University of Waterloo, Waterloo, Ontario, N2L 3G1, Canada\\
        $^4$Dark Cosmology Centre, Niels Bohr Institute, University of Copenhagen, Juliane Maries Vej 30, 2100 Copenhagen, Denmark } 

\begin{document}



\maketitle

\label{firstpage}

\begin{abstract}
The mass aggregation and merger histories of present-day distinct haloes 
selected from the cosmological Millennium Simulations I and II are mapped into 
stellar mass aggregation and galaxy merger histories of central galaxies by using empirical
stellar-to-halo and stellar-to-gas mass relations.  
The growth of bulges driven by the galaxy mergers/interactions is
calculated with dynamical prescripitions. 
The predicted bulge demographics at redshift $z\sim 0$ is consistent
with observations
(\citeauthor{Zavala+2012}). Here we present the evolution of the morphological mix
(traced by the bulge-to-total mass ratio, \bt) as a function of mass up to $z=3$. 
This mix remains qualitatively the same up to $z\sim1$: $\bt\leq0.1$ galaxies
dominate at low masses, $0.1<\bt\leq 0.45$ at intermediate masses, and
$\bt>0.45$ at large masses. At $z>1$, the fractions of disc-dominated and 
bulgeless galaxies increase strongly, and by $z\sim2$ the era of
pure disc galaxies is reached. Bulge-dominated galaxies
acquire such a morphology, and most of their mass, following a downsizing trend.
Since our results are consistent with most of the  
recent observational studies of the morphological mix at different redshifts, a \lcdm-based scenario 
of merger-driven bulge assembly does not seem to face critical 
issues. However, if the stellar-to-halo mass relation evolves too little with redshift, then some tension with 
observations appear.
\end{abstract}


\section{Introduction}
\label{introduction}

The two main stellar components of galaxies are  the disc and the bulge.
The classification of galaxies is tightly related to the luminosity or mass
ratio of these components, for example, to the ratio of bulge
to total (disc+bulge) mass, \bt. Most of the properties of galaxies and 
relevant aspects of their assembly histories are also tightly related to 
this ratio. Thus, a key ingredient in the study of galaxy formation 
and evolution is to understand how is the \bt\ ratio established 
as a function of mass and time. 

According to the general picture of galaxy formation and evolution in the context 
of the $\Lambda$ Cold Dark Matter (\lcdm) hierarchical scenario, discs form generically inside the 
evolving CDM haloes, while bulges grow mainly driven by the merger/interaction of discs 
but also due to intrinsic disc instabilities and by misaligned/perturbed 
infalling gas. The mass accretion and merger histories of CDM haloes as a function of 
mass and environment are calculated precisely by means of N-body cosmological simulations 
\citep[e.g.][]{Lacey+1994, Gottloeber+2001, Wechsler+2002, 
Maulbetsch+2007, Fakhouri+2009,Fakhouri+2010, Zhao+2009, Behroozi+2013-halos}. 
It is common to read in the specialized literature that the predicted \lcdm\ halo merger
rates would imply a large population of galaxies with dominant (merger-driven) classical 
bulges, in conflict with observations \citep[e.g.][]{Weinzirl_2009,Kormendy+2010,Fisher+2011}. 
However, the connection of halo mass assembly and merger rates to the galaxies  they host
is complex and should be properly understood in order to account for the growth of bulges.

Recently, several theoretical studies attempted to establish the connection between the halo merger 
history and the final galaxy \bt\ ratio. Some of these studies are based on the semi-empirical halo 
occupation framework \citep[e.g.][]{Stewart-09,Hopkins-09b,Hopkins-10a},
while others are based on semi-analytic models \citep{Khochfar-06,Parry+2009,Benson+2010,deLucia+2011,
Fontanot+2011}. The most relevant conclusions of these works are that: (i) the mapping of halo-halo 
mergers to stellar galaxy-galaxy mergers is far from linear and strongly depends on mass and 
redshift, (ii) the inclusion of the galaxy gas content in mergers significantly reduces the 
final \bt\ fraction, specially for low-mass galaxies and at high redshifts, 
and (iii) the \bt\ fraction predicted  in the \lcdm~ scenario increases with stellar mass, \ms, in a similar
way as observations, although there seems to be fewer predicted bulgeless galaxies than observed. 

In \citet{Zavala+2012} (hereafter \citetalias{Zavala+2012}),  
the merger-driven bulge formation 
scenario has been revisited by means of a semi-empirical approach
based on \citet[][]{Hopkins-09b,Hopkins-10a}. This approach takes into account:  
(i) the cosmological mass accretion and merger histories of haloes
and an estimate for the time of coalescence of the subhaloes in the centre of the main (distinct) halo;
(ii) the empirically constrained stellar-to-halo and stellar-to-gas mass relations at different $z$; 
(iii) a physically-based model for calculating the stellar and gas mass evolution 
of the satellite galaxies; (iv) dynamical recipes, calibrated with numerical simulations, for calculating 
the mass growth of the bulges after the merger of the primary galaxy with the accreting satellite galaxy. 
The results presented in 
Paper I show that the local bulge demographics and \bt\ vs \ms\ correlation down to galaxies 
of stellar masses $\ms\sim 10^9$ \msun\ are in general consistent with current observational studies. 
It was also shown that the \bt\ ratio depends on the way the satellites evolve until they merge 
with the primary, and that the merger-driven bulges grow in several episodes through
concomitant channels of stellar mass acquisition: from the secondaries, from the primary disc and 
from local starbursts. This produces composite (pseudo + classical) bulges in most of the cases. 

The aim of this paper is to extend the analysis of Paper I by analysing
the evolution of the morphological (\bt\ ratio) mix of galaxies as a function of mass, and to compare our predictions
with the few (recently appearing) observational determinations of the morphological mix at high redshifts.
Some of the questions that we study by means of our semi-empirical model are: 
Does the morphological mix of galaxies strongly change with redshift? What is the epoch of
major changes in this mix? What is the mass dependence of the bulge growth histories of galaxies? 
Does the assembly of early-type galaxies follow a morphological and mass 
downsizing trend? Does the morphological transformation to bulge-dominated galaxies happen 
when galaxies are in their active or passive regime of star formation?

The outline of the paper is as follows. The semi-empirical approach presented in Paper I is 
summarized in Section \ref{model}.  In \S\S \ref{mix}, results on the evolution of the 
\bt\ ratio and the morphological mix as a function of \ms\ are presented, and in \S\S \ref{early-type}
the setting of the bulge-dominated galaxy population is discussed. In Section \ref{observations} we
compare our results with the available observations on morphology and \bt\ ratios at 
different $z$.  In Section \ref{discussion} we discuss the implications of a little evolving 
stellar-to-halo mass relation (SHMR) on the bulge demographics and evolution (\S\S \ref{slowSHMR}), 
and whether the merger-driven bulge growth in the context of the \lcdm\ cosmology is
consistent or not with observations (\S\S \ref{lcdm}). 
Our conclusions are given in Section \ref{conclusions}.

\section{The semi-empirical model}
\label{model}

Our semi-empirical model of galaxy and bulge stellar mass growth is presented in detail 
in Paper I.   Below we summarize its main features.
 
 It is important to remark that we do not model ab initio the physical processes of galaxy evolution (e.g., star 
 formation and feedback), which makes our approach different from semi-analytic models and numerical simulations. 
 Instead, our semi-empirical approach consists in seeding stellar and gaseous masses into the evolving CDM haloes 
 using empirical information. By means of this approach, we can then {\it empirically} extend the halo 
 mass aggregation and merger histories of distinct CDM haloes to the corresponding stellar and gaseous 
 mass aggregation histories of central galaxies, including galaxy merger events. The merger-driven 
 growth of bulges in these semi-empirical galaxies is modeled by using dynamical prescriptions calibrated 
 against numerical simulations.

\subsection{Subhalo merger histories}

Our goal is to analyse the impact of mergers in the growth of the bulges 
of present-day {\it central galaxies}; a central galaxy is defined
to be the most massive galaxy in a given {\it main subhalo}, i.e. a subhalo that is not
contained inside a larger subhalo (known in the literature also as a {\it distinct halo}). 
We therefore extract the merger histories of the principal branches of 
a population of main subhaloes defined at $z=0$.  We have randomly selected two 
samples of such subhaloes, having 1347 and 1500 members with masses larger 
than $1.2\times10^{12}\msun~(10^3~{\rm particles})$ and 
$9.4\times10^{10}\msun~(10^4~{\rm particles})$, from the
Millennium (MS-I) and Millennium II (MS-II) simulations, respectively
\citep{Springel-05,Boylan-Kolchin-09}. 
Both samples (properly normalized to account for the fractional volume they cover relative to
the whole simulation boxes) follow the mass function of the full halo population.
Combining both samples we can follow up to high redshifts  the merger and
accretion histories of haloes having a wide mass range at $z=0$: $10^{10}-10^{15}\msun$. 

A given merger event is characterized by three epochs: (a) the start of the merger, $z_{\rm start}$, 
i.e., when the subhalo was part of an independent friend-of-friends halo for the last time; 
(b) the ``dissolution'' time of the subhalo, $t_{\rm diss}$, i.e., when the merged subhalo at time 
$t_i$ can no longer be resolved  in the simulation as an independent self-bound structure at the following time $t_{i+1}$; 
and (c) the coalescence time of the subhalo  center, where the satellite galaxy is supposed to be,
with the centre of the main subhalo, $t_{\rm end}$.  
To compute the latter we adopt a dynamical friction time formula applied 
just after the subhalo has been dissolved 
 \citep{Binney_1987}:
\begin{eqnarray}\label{merge_time}
t_{\rm df}=\alpha_{\rm fric}(\Theta_{\rm orb})\frac{V_{\rm vir}r_{\rm sub}^2}{G m_{\rm sub}{\ln }~\Lambda},
\end{eqnarray}
where $\alpha_{\rm fric}(\Theta_{\rm orb})$ encloses information on the subhalo orbit, $V_{\rm vir}$ is
the virial velocity of the host, $m_{\rm sub}$ and $r_{\rm sub}$ are the mass and position of the subhalo relative 
to the host just before dissolution, and ${\rm ln}~\Lambda=(1+\mh/m_{\rm sub})$ is the Coulomb
logarithm with \mh\ the virial mass of the host. We take $\alpha_{\rm fric}(\Theta_{\rm orb})=1.17 \eta^{0.78}$
\citep{Boylan-Kolchin-Ma-Quataert-08}, where $\eta=j/j_c(E)$ is the orbital
circularity of the subhalo relative to the halo centre.\footnote{The subhalo 
has specific angular momentum $j$ and energy $E$, and $j_c(E)$ is the
specific angular momentum of a circular orbit with the same energy and with a
radius $r_c(E)$.} 

The final cosmic time of the \textit{halo-halo central merger}, $t_{\rm end}$, 
is the sum of $t_{\rm diss} + t_{\rm df}$;  we consider that $t_{\rm end}$ is a good 
approximation to the actual \textit{galaxy-galaxy merger} epoch.

We note that  the impact of a merger does not depend directly on the halo mass ratio 
at the start of the merger but rather on the central dynamical masses (inner dark matter, 
gas and stars) that interact in the final stages of the galactic merger.

\subsection{Galaxy occupation}
\label{gal-occupation}

To follow the stellar and gas mass assembly of galaxies inside the main haloes, 
as well as the processes that affect the gas and stellar contents during mergers,
we use a semi-empirical approach close to the one in \citet[][]{Hopkins-09b,Hopkins-10a}. 
This approach yields stellar mass assembly histories that are consistent, 
by construction, with observational trends.  The main steps for 
each present-day main subhalo in our MS samples are:

\begin{enumerate}

\item extract the main branch of its merger tree; 

\item seed a central galaxy, the primary, at $z_{\rm seed}\approx 3.5$ with
stellar and gas masses given by the semi-empirical relations 
$M_\ast(M_h,z)$ \citep{Firmani+2010} and $M_g(M_\ast,z)$ \citep{Stewart-09},
see Appendix A for the analytical formulae; 

\item identify the start of a given merger, $z_{\rm start}$, along the main branch 
($z_{\rm start}\leq z_{\rm seed}$) and assign a galaxy to the infalling halo, the 
secondary, according to the semi-empirical relations $M_\ast(M_h,z_{\rm start})$ 
and $M_g(M_\ast,z_{\rm start})$;

\item follow the evolution of the secondary by means of semi-analytic recipes 
assuming that the satellite galaxy does not accrete more gas (moderate quenching); 
for this, its gas mass is distributed in an exponential disc that transforms gas
into stars with a local Kennicutt-Schmidt law and ejects gas at each radius 
due to SN feedback in form of energy-driven outflows (for details see Appendix
A3 of Paper I); 

\item compute the galaxy-galaxy (central halo) merging time, $t_{\rm end} = t_{\rm dis}
+ t_{\rm df}$;

\item estimate the bulge (and disc) masses of the primary galaxy after coalescence at $t_{\rm end}$, 
using physical recipes for the bulge growth calibrated by numerical simulations (see \S\S \ref{bulge_growth} below);

\item repeat items ii--vi 
until reaching $z=0$, taking care of each merger, the bulge growth, and updating at each $z$ the properties of the central
galaxy according to the $M_\ast(M_h,z)$ and $M_g(M_\ast,z)$ relations.

\end{enumerate}

It is important to remark that in our scheme, after a major merger happens, and the bulge has
formed, the mock galaxy will likely continue growing (it depends on the halo mass assembly history and
the $M_\ast(M_h,z)$ and $M_g(M_\ast,z)$ relations). We assume that this ``smooth" growth 
corresponds to the disc only. Once a new merger happens, the bulge may then 
grow according to item vi. This is supported by numerical simulations, whence
several authors have shown that even after a major merger, the disc may regenerate from the available gas and/or rebuild
by late gas accretion  (for theoretical works see e.g., \citealp{Robertson+2006, Governato+2009,Hopkins-09a,Tutukov+2011},
and for observational evidence see \citealp{Hammer+2009,Puech+2012}).

\subsubsection{Initial conditions}
\label{initial-conditions}

The empirical relations we use are poorly constrained at $z>3$ and the mass aggregation and 
merger histories of the less massive haloes are not reliable for these epochs due to resolution
issues. We therefore choose an initial redshift $z_{\rm seed}\sim 3.5$, where we ought to define
an initial condition for the \bt\ ratio of the seeded galaxies.  
In Paper I, we assumed that all galaxies are initially pure discs. The galaxy population obtained
at $z\sim 0$ is essentially independent of this initial condition because haloes and galaxies grow 
significantly after more than 10 Gyr of evolution. The \bt\ ratio of even the most massive 
bulge-dominated galaxies is mainly stablished at $z\sim 2-1$ due to the major merger activity 
these galaxies suffer at these epochs (see \S\S 3.2 below). 

Because our aim here is to study the morphological mix up to $z\sim 3$, 
and make comparisons with observations, the initial conditions need to be treated more carefully than in Paper I since
it influences the results near this epoch, at least down to $z\sim 1.5$.  
Thus, we assume that if a given halo at $z_{\rm seed}$ has a mass larger than the corresponding to a 2-$\sigma$ halo,\footnote{
A collapsed structure of mass $M$ is said to have a peak height $\nu=\delta_c^2(z_{\rm coll})/\sigma^2(M)$,
where $\sigma$ is the linear mass variance and $\delta_c$ is the critical overdensity required for 
spherical collapse.  
A 2--$\sigma$ halo of mass $M$ has $\nu=2$ at $z=z_{\rm coll}$.} then its 
central galaxy is seeded with \bt=0.9; otherwise, the galaxy is assumed 
to start as a pure disc. This implies that only the most massive haloes at $z_{\rm seed}$
host bulge-dominated galaxies at that epoch. 

Both theory and observations indicate that most galaxies should be indeed disc-dominated 
at redshifts as high as 3--4. For example, the gas fraction-dependent model of Hopkins et al. (2009b) 
shows a much higher fraction of disk galaxies ($\bt < 0.25$) at different masses compared to 
bulge-dominated galaxies ($\bt > 0.7$) at $z=3$.  
Hydrodynamical simulations constrained to reproduce today a massive spiral galaxy 
\citep{Guedes+2013} and a massive spheroidal galaxy \citep{Naab+2009}, 
show a disk-like structure (with S\'ersic index $n < 2.5$) at $z\sim 3$ in both cases. 
Regarding observations at $z>2$, several studies of morphology, \bt\ ratio, S\'ersic index, etc. 
show that disc, late-type galaxies are clearly more abundant,
even for massive galaxies \citep{Bruce+2012,Buitrago+2013,Mortlock+2013}. 
In Section 4.2 we discuss in detail many of the results from these works. 

Nevertheless, at high redshifts there is also a (small) fraction of massive galaxies that are already 
bulge-dominated (ellipticals) or are in the process of becoming bulge-dominated (sub-millimeter
galaxies, SMGs).  According to observational studies, the massive, old ellipticals are likely  
descendants of SMGs \citep[e.g.,][]{Hickox+2012,Riechers+2013, Toft+2014}, being both 
associated to massive, clustered, high-$\sigma$ haloes.  Haloes with higher $\nu$ values (more massive
than the average) collapsed earlier than those with lower values, and thus, they have been subjected to a larger 
number of major mergers \citep[e.g.,][]{Lagos+2009}. High-$\sigma$ haloes are associated to high peaks in the density 
fluctuation field, and high peaks are clustered and surrounded by other peaks, i.e., the whole 
region has a higher density and collapses earlier than the average region for that scale
\citep{Bardeen+1986, Bond+1991}.  
As it will be clear in Section 4.2, the observed fraction of massive bulge-dominated galaxies 
at $z=2-3$ is nearly reproduced with our assumption of haloes with $\nu\ge 2$ at $z_{\rm seed}\sim 3.5$
hosting galaxies with $\bt=0.9$. 
We emphasize once again that for most of our results, the initial conditions are already not relevant for $z\lesssim1.5$.

\subsubsection{Satellite galaxies and merger rates}

Regarding the fate of galaxies once they become satellites, the semi-analytical model used to
follow them implies that their stellar mass growth is only slightly less healthy than the one 
of centrals of the same mass (Paper I). Semi-empirical studies seem to confirm this behavior 
\citep{Watson+2013,Rodriguez-Puebla+2013,Wetzel+2013}. In Paper I we have explored also the
case of extreme satellite quenching (the stellar and gas masses of the satellites do not
change since the  time of accretion), and obtained an unrealistically low fraction of classical-like bulges.

We highlight that, as shown in Paper I, the stellar major merger rates as a function 
of $z$ obtained with our \lcdm-based semi-empirical fiducial model are in good agreement
with several observational inferences \citep[see also][]{Hopkins-10a, Puech+2012}. Here we use 
also this \textit{fiducial} model, which corresponds to the stellar-to-halo mass relation (hereafter SHMR), $M_\ast(M_h,z)$,  
given in \citet{Firmani+2010} and the evolving satellite case described above. 
In \S\S \ref{slowSHMR} we discuss the effects on our results when varying the SHMR evolution.

\subsection{Bulge growth channels}
\label{bulge_growth}

The bulge growth in our scheme is driven by galaxy mergers, but this growth is not 
limited only to the acquisition of stars from the secondaries. 
The three channels of stellar mass bulge growth used in item (vi) of the previous section are: 

\begin{itemize}

\item (a) incorporation of all the stars of the merged secondary;

\item (b) disc instability-driven transport of a fraction of 
stars in the primary galaxy, $f_{\rm relaxed}^p$;

\item (c) newly formed stars in central starbursts produced by a
fraction of the gas from both merging galaxies, $f_{\rm burst}^{\rm p+s}$.

\end{itemize}

In the last stages of coalescence, the surviving dynamical mass of the secondary $M_2$  
collides with the central region of the primary, of dynamical mass $M_1$. The dynamical
mass is defined as the sum of dark, stellar, and gas masses inside the halo
scale radius, $r_s$, where the halo mass distribution is approximated by the NFW
density profile \citep{NFW_1997}. Thus, $r_s=r_{\rm vir}/c$,
where $c$ is the halo concentration; the c(\mh,$z$) relation 
of \citet{Gao-08} is used. During  coalescence,
the collisionless components of both systems are subject to rapid changes
of the gravitational potential that broaden their energy distributions leading towards an equilibrium
state. This relaxation process drives the stars originally rotating in discs towards random 
orbits forming a spheroidal remnant. A simple dynamical argument shows that the stars in 
the primary disc affected by the action of the secondary are those within a radius enclosing 
the mass corresponding to $\sim M_2$; the stars at larger radii in the disc are also perturbed but likely
they are re-arranged into final configurations that are not far from the original ones.
Thus, the fraction of the primary stellar disc that relax into the central spheroid, $f_{\rm relaxed}^p$,
is roughly given by the dynamical mass ratio, $\mu_{\rm eff}\equiv M_2/M_1$.  A large set 
of numerical simulations performed by \citet{Hopkins-09a} confirm this approximation, but 
in more detail, they suggest a slightly non-linear dependence on $\mu_{\rm eff}$:
$f_{\rm relaxed}^p\approx \mu_{\rm eff}\times 2(1 + \mu_{\rm eff}^{-a})^{-1}$, 
with $a=0.3-0.6$ \citep{Hopkins-09b}. We adopt this correction and use $a=0.3$ for our fiducial model

During final coalescence, the interaction generates also a non-axisymmetric response in the 
galactic discs that morphologically resembles a bar. The resulting stellar and gaseous bars are 
however out of phase because gas is collisional and stars are not. Because of this, the stellar 
bar torques the gas bar draining its angular momentum. In this way, the cold gas is effectively 
removed from the original discs and transformed into stars, during a starburst, in the bulge 
of the remnant. This process is efficient within a region inside a critical radius $r_{\rm crit}$,
which depends on the merger mass ratio and relative orientation and orbit of the progenitors, as
well as their stellar and gaseous content. A parametrization of this ratio, $f_{\rm burst}^{\rm p+s}$,
obtained from numerical simulations, is given in \citet[][see also Hopkins et al. 2009a]{Hopkins-09b};
we use this parametrization to calculate $f_{\rm burst}^{\rm p+s}$ (see details in Paper I).

As showed in Paper I, bulges are composite, i.e., their stars are acquired
by the three channels. However, for massive galaxies, $\ms>10^{11}$ \msun, 
channel (a) dominates; for smaller galaxies, channel (b) dominates;
and channel (c) contributes with only a minor fraction  ($<10\%$) of the bulge 
mass in all the cases. Note that we do not account for intrinsic secular disc
instabilities, which typically are associated with the formation of pseudo-bulges.
However, our channel (b) can be associated also with the formation 
of pseudo-bulges (stars come from the same disc), 
while channel (a) will likely give rise to a classical bulge.
 
The dominion of a given channel as a function of \ms\ is closely related to the merger 
history of individual galaxies and their gas fractions. 
Most of the massive galaxies assembled a significant fraction 
of their stellar masses by major stellar mergers with small gas fractions, which leads to 
the growth of prominent bulges dominated by stars from the secondaries 
(see middle panels of Fig. 3 in Paper I). For
less massive galaxies, minor/minuscule stellar mergers with high gas fractions dominate,
which leads to (small) bulges formed mainly from dynamically 
perturbed stars of the primary disc (the stellar mass merger ratio is very small but the
dynamical mass merger ratio -due to the high gas fraction- is large enough 
to produce the disc instability; see right panels of Fig. 3 in Paper I).

\section{Results}	
\label{results}

We now present the results of the semi-empirical approach described above for the
case of the fiducial model. We recall that: (i) the analysed
mock galaxies have been defined at $z=0$ to be in distinct haloes; therefore these galaxies are
centrals; and (ii) the evolutionary trends shown below refer to the evolution of these present-day
central galaxies. If a halo (galaxy) is distinct (central) at $z=0$, then the it is very likely that it was
distinct (central) in the past as well.\footnote{The exception are the ``backsplash'' haloes at $z=0$, i.e., 
those that passed through larger haloes in the past but today are outside appearing as 
distinct haloes. They are a small fraction, $4-9\%$ according to \citet{Wang+2009},
with the fraction being increasingly smaller for larger masses.}
Thus, we are confident that our results refer mostly to distinct haloes and central galaxies at 
{\it all} redshifts, and that the populations of our ``evolved" central galaxies at different redshifts 
describe well the overall population of central galaxies at a given redshift.
 For example, we have checked that the Galaxy Stellar Mass Function at $z=0$ 
agrees well with the measured one from observations.

\begin{figure}
\centering
\includegraphics[height=10cm,width=8.8cm,trim=0.2cm 0.2cm 0.2cm -0.cm, clip=true]{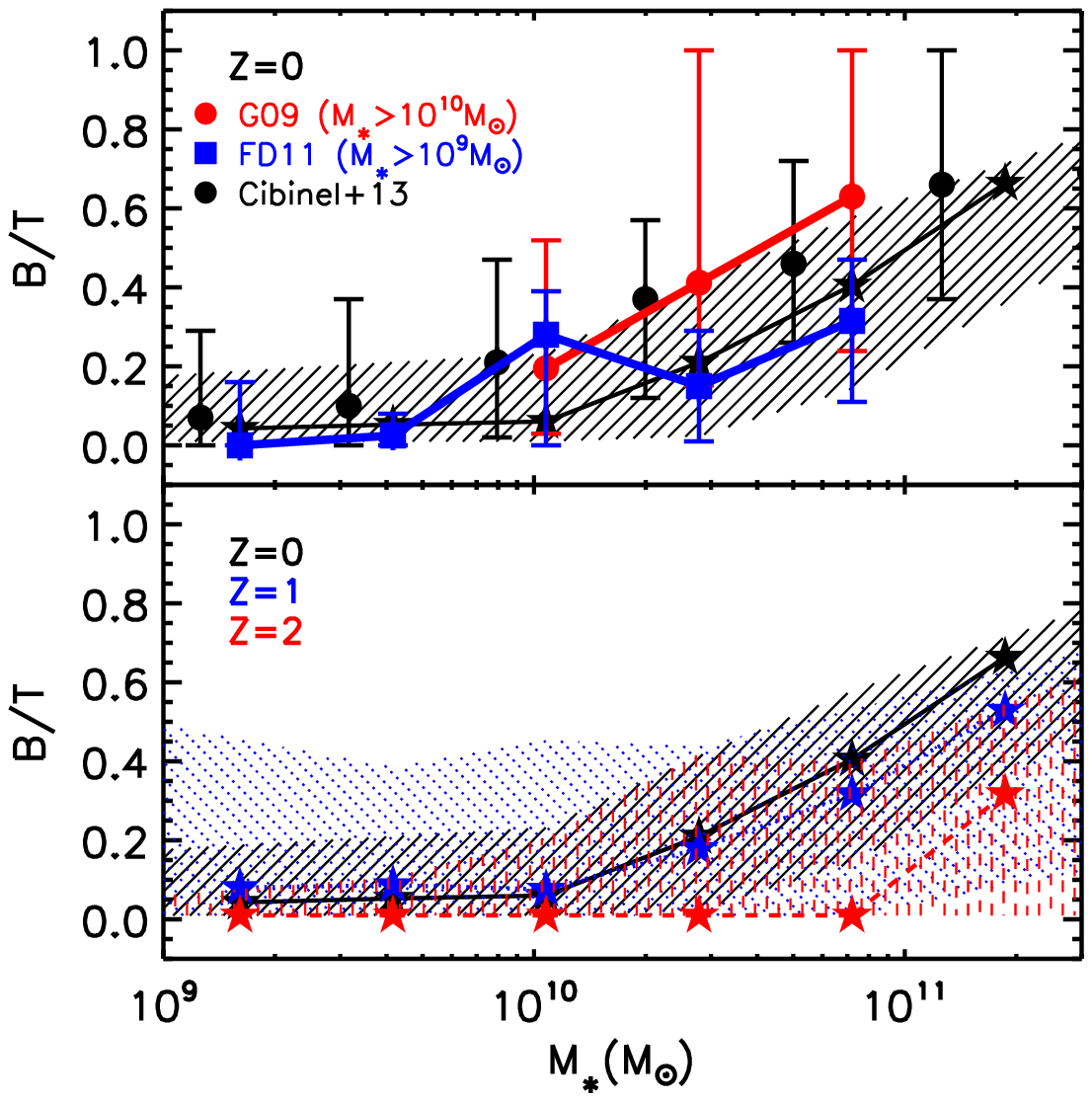}
\caption{Distributions of the \bt\ ratio as a function of stellar mass. Upper panel: the median and $1\sigma$ scatter
of the semi-empirical galaxies (stars connected by the black line and the shaded area), the means of the 
observed sample of galaxies from 
\citet[][dots with error bars]{Cibinel+2012}, and the means of the 11 Mpc-volume sample of 
\citet[][squares connected by the blue line]{Fisher+2011} and the volume-complete 
sample of \citet[][dots connected by the red line]{Gadotti2009}. 
Lower panel: as in the upper panel for the semi-empirical galaxies but at three
redshifts: $z\sim 0$ (black solid line and diagonal-line shaded area), $z\sim 1$ 
(blue dotted line and dotted shaded area),
and $z\sim 2$ (red dashed line and vertical-line shaded area).
}
\label{BTvsMs-local}
\end{figure}

\subsection{Evolution of the morphological mix as a function of mass} 
\label{mix}

\begin{figure*}
\centering
\includegraphics[height=5.5cm,width=5.5cm,trim=1.2cm 0.5cm 1.0cm 1.0cm, clip=true]{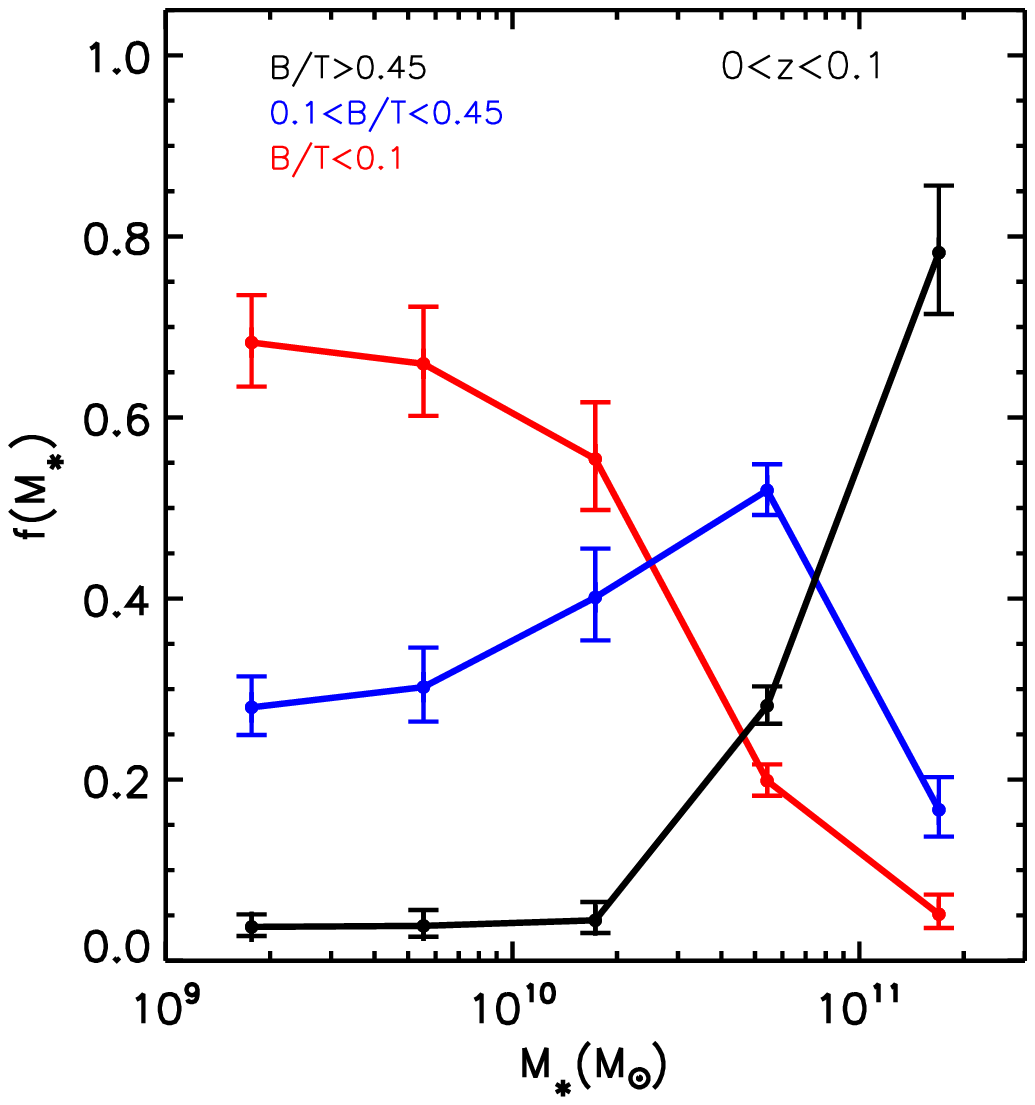}
\includegraphics[height=5.5cm,width=5.5cm,trim=1.2cm 0.5cm 1.0cm 1.0cm, clip=true]{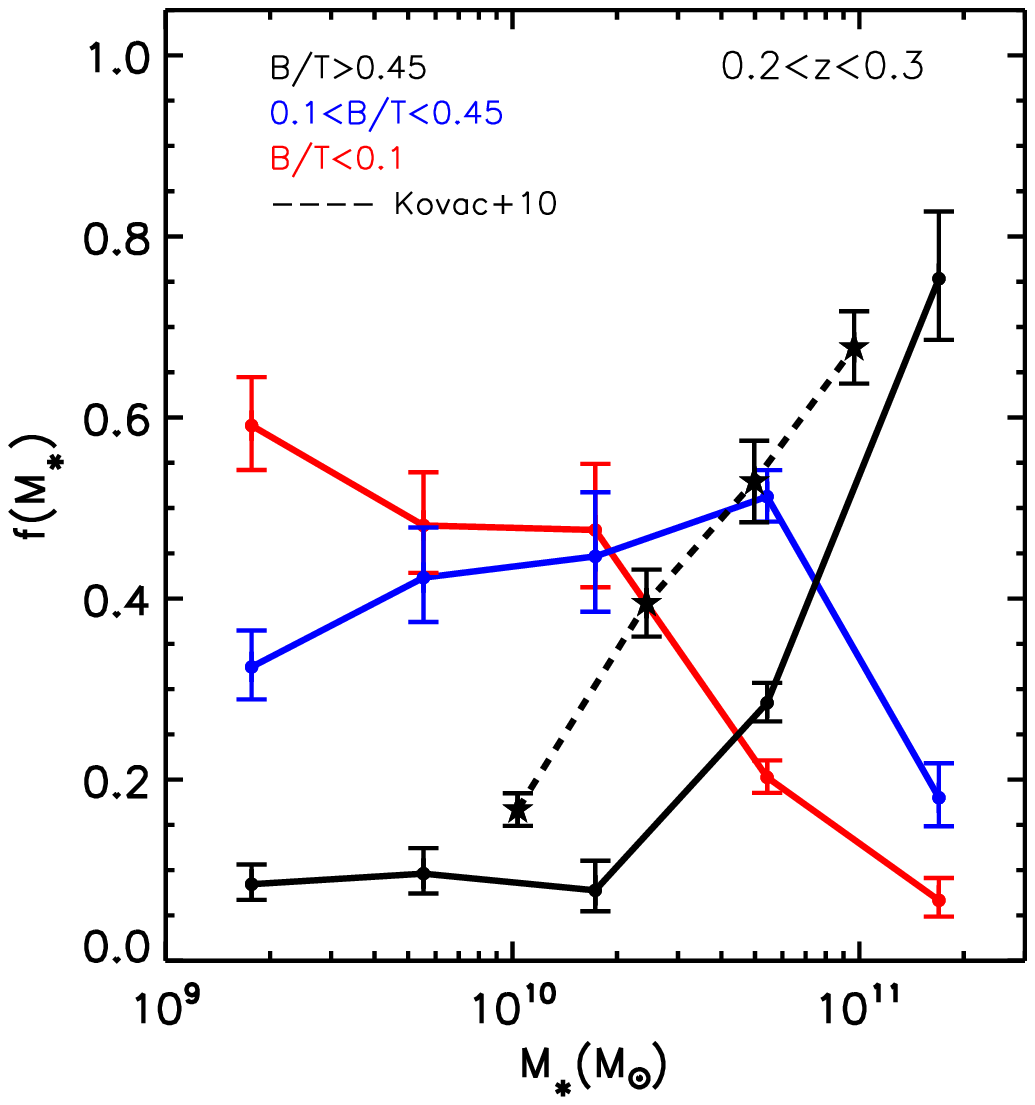}
\includegraphics[height=5.5cm,width=5.5cm,trim=1.2cm 0.5cm 1.0cm 1.0cm, clip=true]{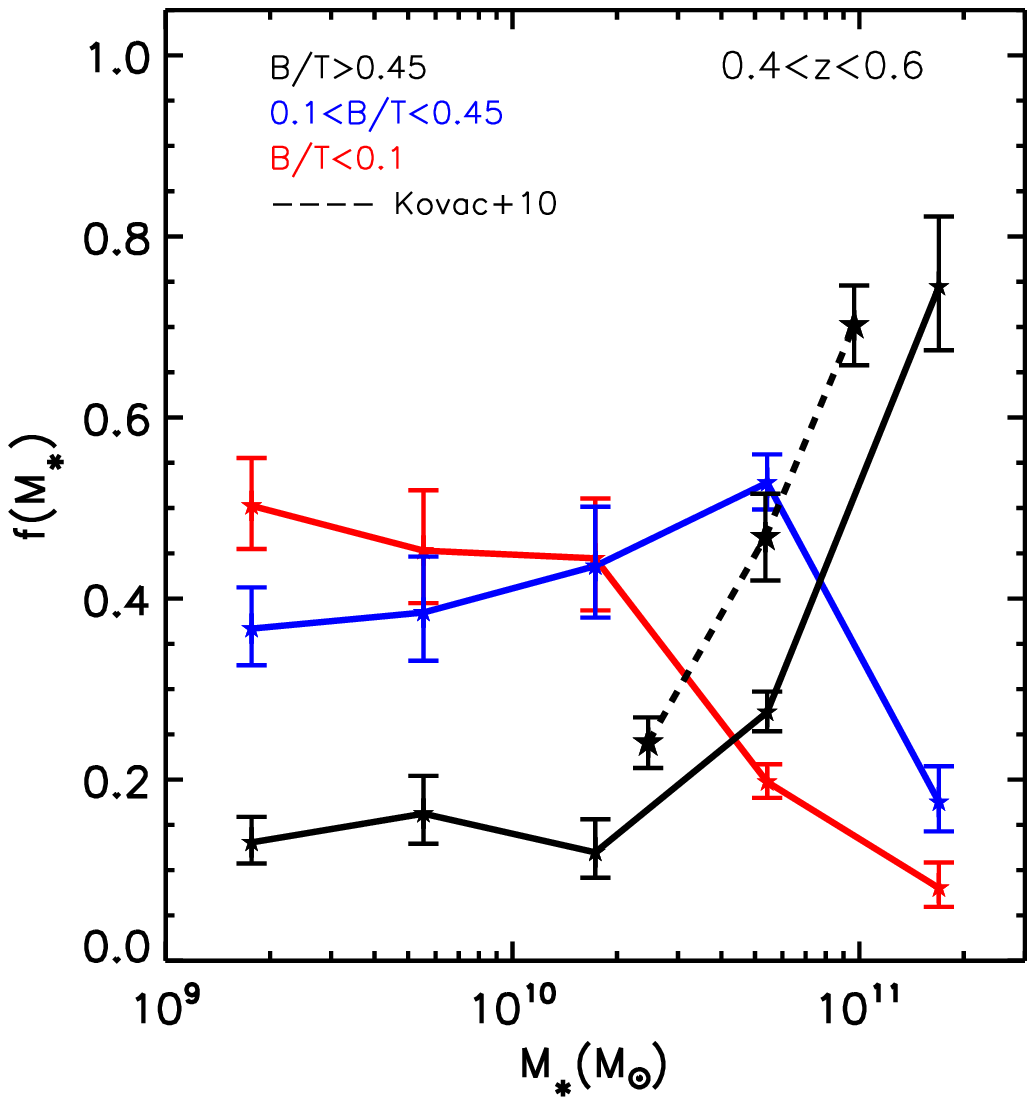}
\includegraphics[height=5.5cm,width=5.5cm,trim=1.2cm 0.5cm 1.0cm 1.0cm, clip=true]{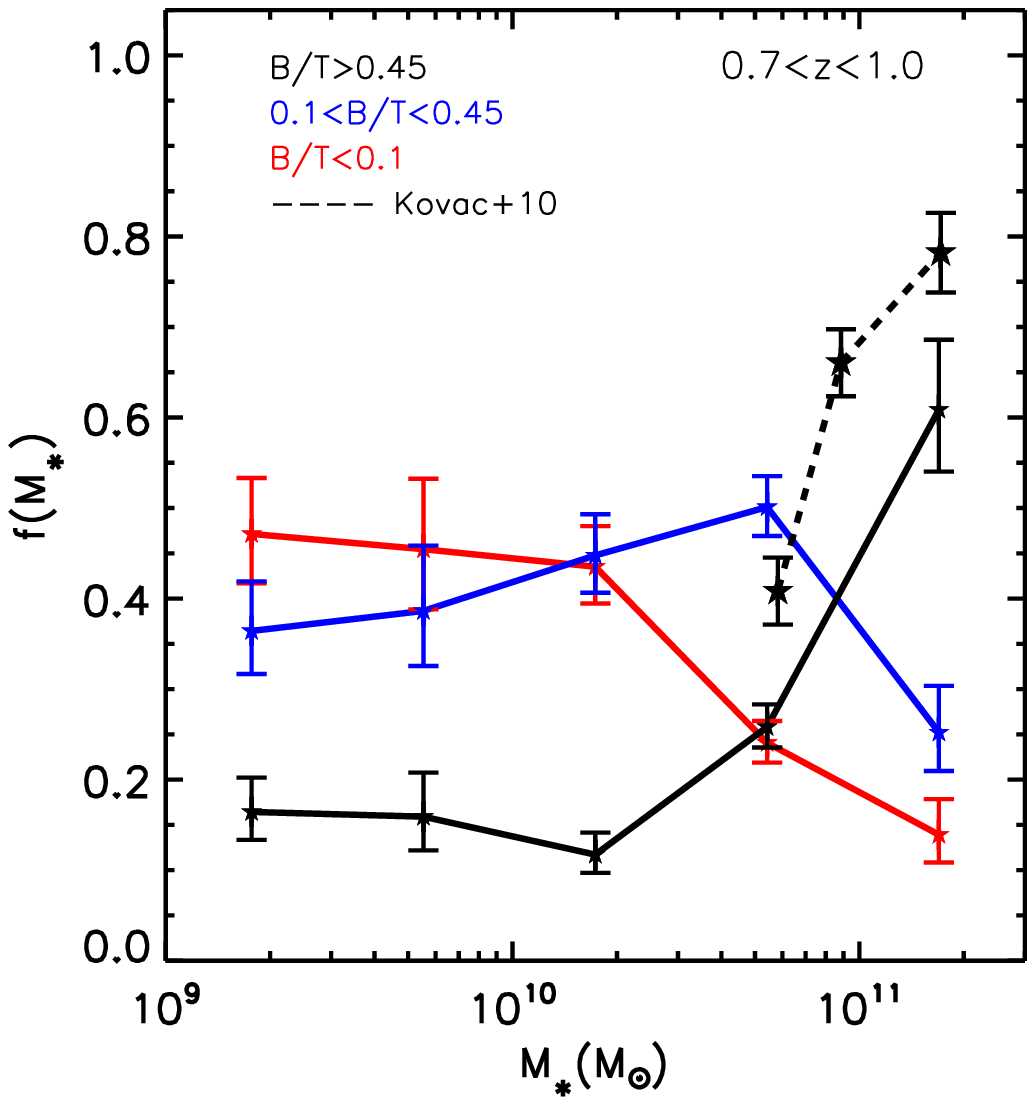}
\includegraphics[height=5.5cm,width=5.5cm,trim=1.2cm 0.5cm 1.0cm 1.0cm, clip=true]{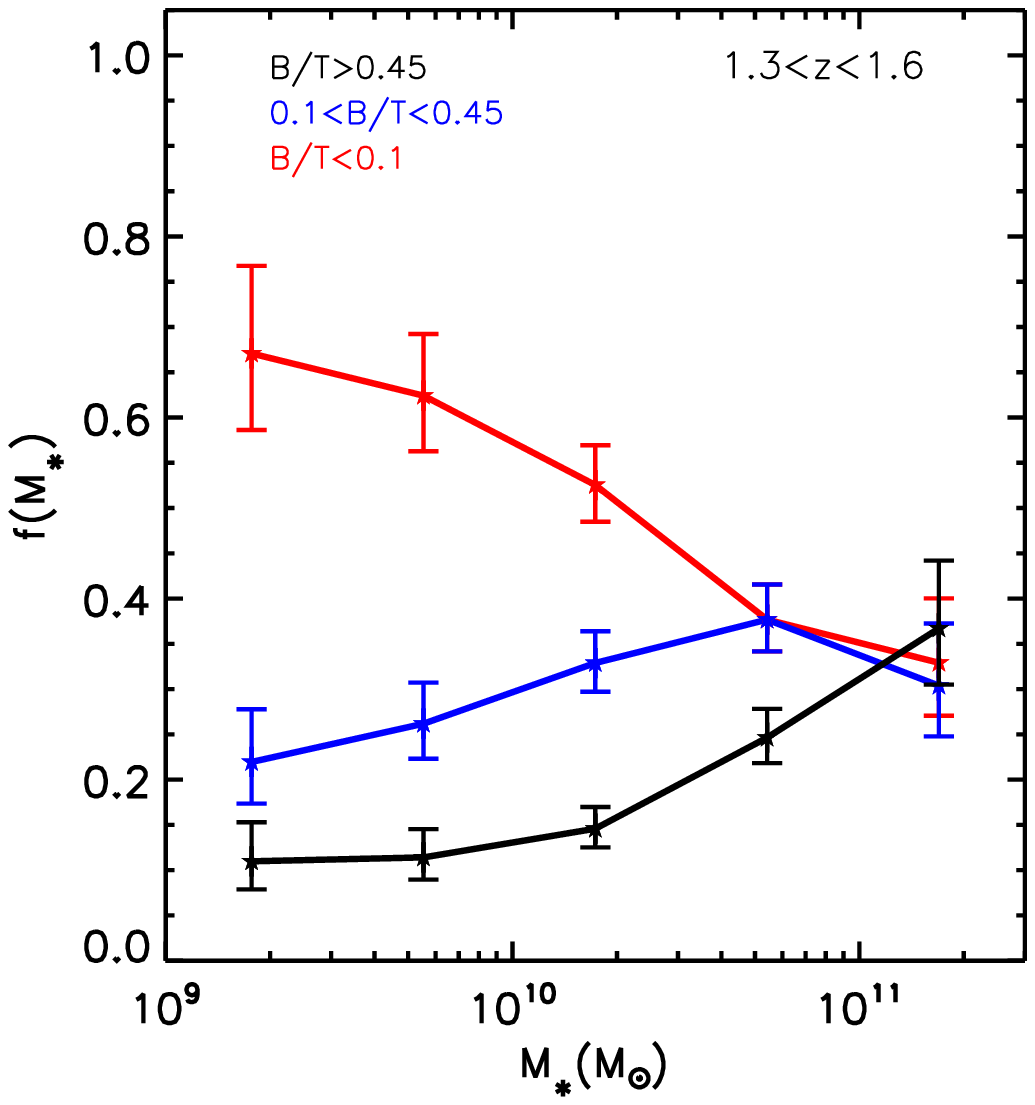}
\includegraphics[height=5.5cm,width=5.5cm,trim=1.2cm 0.5cm 1.0cm 1.0cm, clip=true]{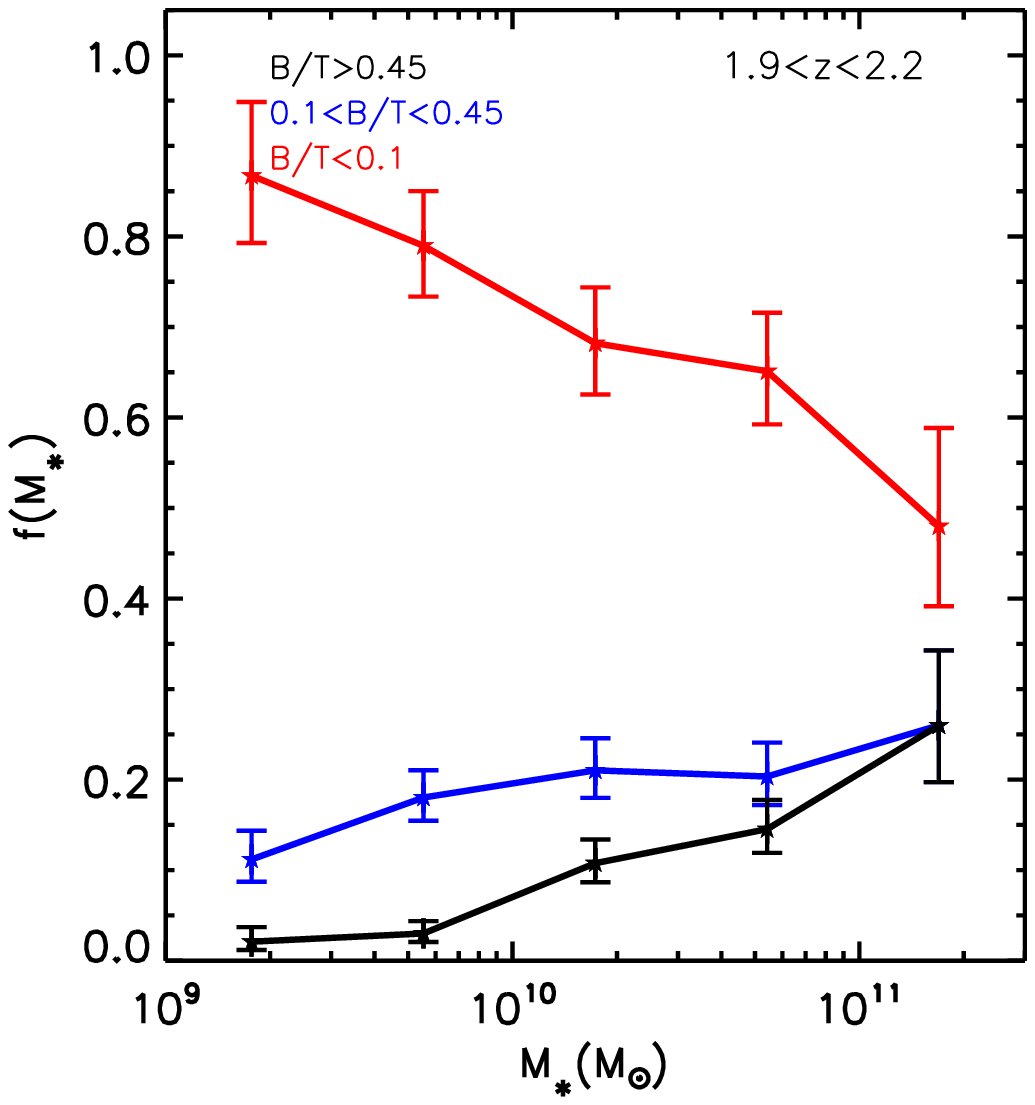}
\caption{Predicted evolution of the relative fractions of galaxies as a function of \ms\ according to their \bt\ ratio: 
$\bt \leq 0.1$ (red dots), $0.1 < \bt\leq 0.45$ (blue dots) and $\bt> 0.45$ (black dots). From left to right and top to bottom,
the fractions correspond to redshift intervals that increase from $z\sim 0$ to $z\sim 2$; the intervals are shown in each panel.
The stars with error bars in some panels correspond to the fractions of E+B galaxies measured in the $z$COSMOS survey
by \citet[][see \S\S \ref{high-z} for details]{Kovac+2010}.
}
\label{BT-histograms}
\end{figure*}

A notable prediction of the semi-empirical model is the strong dependence of the \bt\ ratio on stellar mass
at $z=0$, with central galaxies smaller than $\ms\sim 10^{10}$ \msun\ having typically $\bt<0.2$, and 
larger central galaxies having higher \bt\ values as \ms\ increases (Paper I; see also \citealp{Hopkins-09b}). 
The median \bt\ ratio versus \ms\ and the 1$\sigma$ regions of the distribution are plotted for $z=0$ galaxies in the  upper panel of 
Fig. \ref{BTvsMs-local} (black solid line and diagonal-line dashed area). In this figure we also plot observational 
 inferences that will be discussed in Section \ref{observations}. In the lower panel the same plot is repeated 
and we also show the model data at $z=1$ (blue dotted line and shaded area) and $z=2$ (red dashed line 
and vertical dashed area). We can appreciate two main results regarding the evolution of the \bt--\ms\ relation:
(i) From $z=0$ to $z=1$, the average of the \bt--\ms\ relation almost does not change, while
from $z=1$ to $z=2$ a significant reduction of the \bt\ ratio is observed at all masses. (ii) The scatter in
the \bt\ distribution is lower at $z=0$ than at higher redshifts, showing that \textit{the morphology of central galaxies, as traced 
by the \bt\ ratio, becomes better defined at later epochs.} 
 At a given epoch, for a given \ms, the source of scatter in Fig.~\ref{BTvsMs-local} is the 
stochastic nature of the prior stellar/baryonic merger history of galaxies. 

Figure \ref{BT-histograms} shows the fractional distributions as a function of \ms\ of the semi-empirical galaxies
with \bt\ ratios $<0.1$ (red line), between 0.1 and 0.45 (blue line) and $>0.45$ (black line). This plot can be 
interpreted as the morphological mix of galaxies if the \bt\ ratio is assumed to be a good morphology indicator.
The distributions are shown at 6 redshift bins: 0-0.1, 0.2-0.3, 0.4-0.6, 0.7-1.0, 1.3-1.6, 1.9-2.2 , 
from top-left to bottom-right, respectively. At $z\sim 0$, ``bulgeless" (\bt$<0.1$) galaxies are the most 
frequent for log(\ms/\msun)~$\lesssim 10.3$, disc-dominated galaxies ($0.1\leq\bt<0.45$) are the most 
frequent for 10.3~$\lesssim$log(\ms/\msun)$\lesssim 10.9$, 
and for larger masses, the bulge-dominated galaxies ($\bt\geq0.45$) are already the most frequent. 
\textit{For redshifts up to $z\sim 1$, the morphological mix remains qualitatively the same}.
In more detail, from $z=1$ to $z=0$, there are two quantitative differences in the fractional distributions:
an increase of bulgeless galaxies at low masses (see below for an explanation of this result),
and an increase of galaxies with $\bt>0.45$ at very high masses.
A strong change in the morphological mix is
observed at redshifts higher than $z\sim 1$: bulgeless galaxies highly dominate 
and the bulge-dominated galaxies become rare, even at the largest masses. Since $z\sim 1.5$, the
morphological mix is already qualitatively different with respect to lower redshifts. At $z\sim 2$, 
73\% of all galaxies more massive than log(\ms/\msun)=9 have $\bt<0.1$, while only $\approx 25\%$ 
of the galaxies more massive than log(\ms/\msun)=11 are bulge-dominated.  

\begin{figure}
\centering
\includegraphics[height=10.3cm,width=8.7cm,trim=0.5cm -0.2cm 0.5cm -0.cm, clip=true]{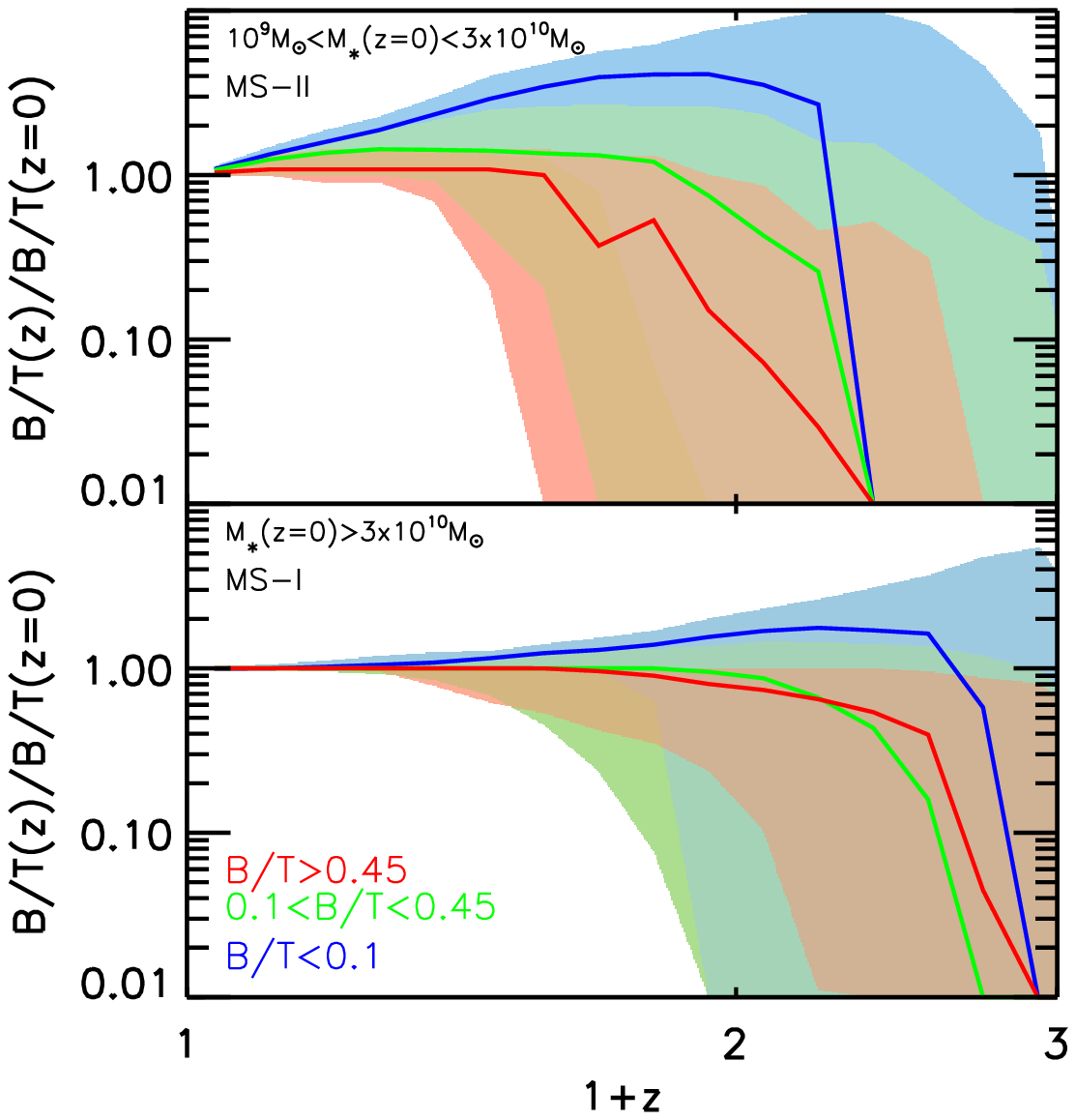}
\caption{Evolution of the \bt\ ratio normalized to the present-day value. 
The  medians and $1\sigma$ regions of the distribution of the semi-empirical
galaxies grouped in three samples according to their \bt($z=0$) value are plotted: blue line and shaded area,
green line and shaded area, and red line and shaded area are for \bt($z=0$) lower than 0.1, in between 0.1 and 0.45, 
and higher than 0.45, respectively. The upper (lower) panel is for present-day central galaxies smaller (larger) than 
$\ms=3\times 10^{10}$ \msun.}
\label{BTevolution}
\end{figure}

In Fig. \ref{BTevolution}, we plot the evolution of the \bt\ ratio of our semi-empirical galaxies, normalized
to their present-day value. The galaxies are divided into three groups according to their 
present-day \bt\ value (as indicated in the figure), and the corresponding median and 
$1\sigma$ regions of the distribution for each $z$ are plotted. The upper and lower panel are for galaxies less and 
more massive than \ms$=3\times 10^{10}$ \msun, respectively. 

For the less massive galaxies (upper panel), most of those
that end today with $\bt<0.1$ (the majority) had larger \bt\ ratios in the past (up to $z\sim 1$);
this explains what is observed in Figs. \ref{BTvsMs-local} and \ref{BT-histograms}.  
These galaxies have formed a merger-induced bulge by $z\sim 1$, after that their 
stellar masses kept growing in the smooth (no merger) regime in such a way that their discs 
grow, making their \bt\ ratios very small by $z\sim 0$. The very few low-mass galaxies 
that today have larger \bt\ ratios, e.g. $\bt>0.45$,
assembled their bulges later on average ($z\sim 0.5$), with almost no change in their morphology since then.
For the massive galaxies (lower panel), their \bt\ ratio becomes defined on average earlier than
for less massive galaxies; the trends of the \bt\ ratio evolution for the different present-day morphologies
are similar to those described above for the low-mass galaxies, but much weaker. 
The massive bulge-dominated galaxies ($\bt>0.45$ at $z=0$) acquired their morphology
between $z\sim 1.5$ and 0.5, and since then, their \bt\ ratios have increased very little.

\subsection{The setting of the bulge-dominated galaxy population}
\label{early-type}
  
\begin{figure}
\centering
\includegraphics[height=7.5cm,width=8.5cm,trim=1.2cm 0.5cm 0.5cm -0.5cm, clip=true]{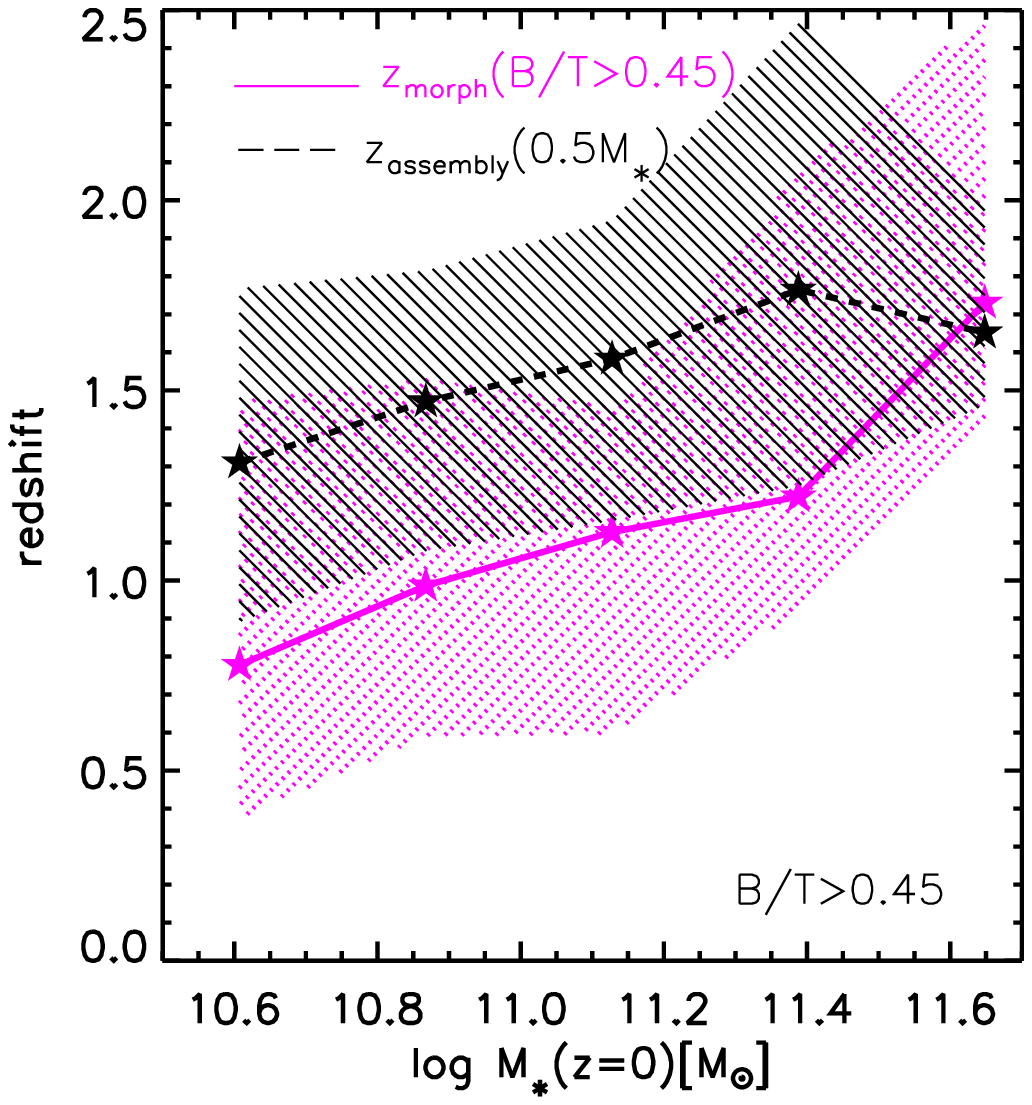}
\caption{The  median redshift and the $1\sigma$ region of the distribution at which present-day 
bulge-dominated semi-empirical galaxies of mass \ms\
attained this morphology (magenta solid line and dot-shaded area) and 50\% of this mass (black dashed
line and diagonal-line shaded area). Galaxies clearly follow a morphological and mass downsizing trend.
}
\label{zmorph-Ms}
\end{figure}

In Fig. \ref{zmorph-Ms} we plot the median of \zmorph, the redshift at which a present-day bulge-dominated 
(early-type) central galaxy attained a \bt\ value larger than 0.45 (solid line), i.e. when it became of early-type; 
the dashed region shows the $1\sigma$ region of the distribution. 
The plot shows that {\it the less massive the present-day bulge-dominated galaxy is, the later it attained such a 
morphology}. The scatter in \zmorph\ increases for less massive galaxies. However, recall 
that the fraction of $z=0$ low-mass bulge-dominated galaxies is very small (Paper I), thus, the large scatter
at $\ms\lesssim 7\times 10^{10}$ \msun\ could be just due to low-number statistics.    
We also plot the median redshift, \zass, at which 50\% of the present-day bulge-dominated galaxy 
has been dynamically assembled (black dashed line); the black shaded area shows the $1\sigma$ region of the distribution. 
From Fig. \ref{zmorph-Ms} we can see that \zmorph\ and \zass\ are closely related, which implies that
the dynamical mass assembly of the $z=0$ bulge-dominated galaxies is driven by major mergers 
(see also Paper I), and their merger histories are such that, on average, the larger the galaxy, the earlier
 it suffered the last major stellar merger that transformed it into a bulge-dominated one \citep[see also][]{Hopkins-09b}.

Our results show that present-day early-type central galaxies assembled under a \lcdm\
merger-driven scenario, {\it follow mass assembly and morphology downsizing trends}, i.e.,
the less massive the early-type galaxy, the later it assembled half its stellar mass and  
the later it became bulge dominated. Besides, on average, these galaxies transit to bulge-dominated 
after they attained half their masses, except the most massive ones. 
 The mass assembly downsizing is just a consequence of the halo mass aggregation histories and the 
empirical SHMRs we have used. 
From this combination, the larger the present-day \ms\ is for a given galaxy, the earlier it
assembled most of its mass \citep[see e.g.,][]{Firmani+2010, Behroozi+2013-main}. The possible
physical explanation behind this is that massive galaxies assembled most of their masses by
early efficient wet mergers, thus, their growth is slowed down, in spite that their host haloes continue growing.
This is because of (i) the long radiative cooling time of the gas in haloes with high circular 
velocities, and (ii) the efficiency of AGN feedback for massive galaxies \citep[e.g.,][]{Croton+2006,Cook+2009}.

Several independent 
observational pieces of evidence show that early-type galaxies follow indeed a mass assembly 
downsizing \citep[c.f.][]{Cimatti+2006, Thomas+2010, Pozzetti+2010}, although probably stronger
than in our case. Semi-analyitic models obtain an opposite behavior for the
dynamical mass assembly of early-type galaxies (for a comparison of several models, see Fig. 
18 in \citealp{Pozzetti+2010}). This finding has been used as an argument against the \lcdm\ scenario. 

\begin{figure}
\centering
\includegraphics[height=7.5cm,width=8.5cm,trim=1.2cm 0.5cm 0.5cm -0.5cm, clip=true]{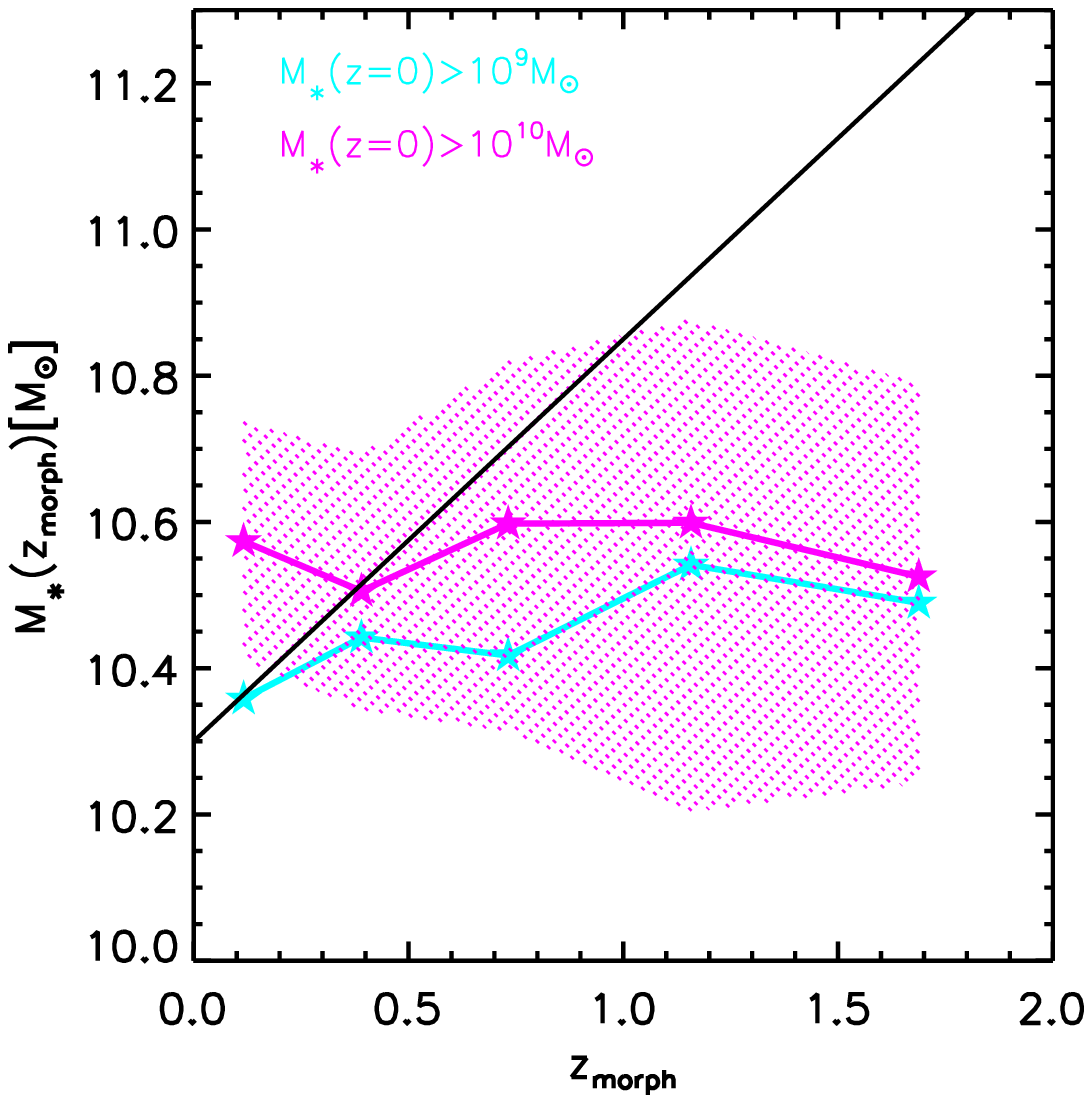}
\caption{The median mass and the first and third quartiles of the distribution of the semi-empirical galaxies 
that are transiting to bulge-dominated  ($\bt>0.45$) at a given $z$. The magenta (cyan) line and shaded 
area correspond to galaxies with \ms\ ($z=0$) $>10^{10}(10^9)$M$_\odot$. 
The black solid line is the fit given in \citet{Firmani+2010} to the 
mean mass at a given $z$ that is transiting from active to passive stellar mass growth (it  also applies to our semi-empirical
galaxies); galaxies of masses above (below) this curve are in their passive (active) mass growth regime. 
For $z>0.5$, the semi-empirical galaxies transform into bulge-dominated typically when they are still active. 
For  $z<0.5$, this transformation typically happens when the galaxies are passive. 
}
\label{Mtrans-z}
\end{figure}

Is there a typical mass for those central galaxies that are transiting to bulge-dominated 
systems ($\bt>0.45$) at a \textit{given epoch}? 
The magenta (cyan) stars joined by a solid line in Fig. \ref{Mtrans-z} show the median of the distribution
of stellar masses of galaxies making this morphological transition at a given redshift bin. 
The shaded area brackets the first and third quartiles of the distribution. Have in mind that this plot
takes into account the number of galaxies at each mass.
Since most of bulge-dominated galaxies are massive but massive galaxies are not dominant in number,
we consider only those with \ms($z=0$) $>10^{10}$ \msun, although for completeness, we plot also 
the median for the case \ms($z=0$) $>10^{9}$ \msun, cyan solid line.  The current morphological 
transition mass does not vary significantly with $z$. The scatter in the distribution of this mass 
at each $z$ is clearly larger than the possible change with $z$. 
Note  that, despite including all galaxies with \ms\ ($z=0$) $>10^9$ \msun, the median morphology 
transition mass is at all $z$ larger than $\ms\sim 2.5\times 10^{10}$ \msun, in 
agreement with the fact that most of low-mass galaxies were never bulge dominated.

The black solid line in Fig. \ref{Mtrans-z} corresponds to the current ``quenching" transition mass given in 
\citet{Firmani+2010}, based on the connection of the semi-empirical SHMR relations at different $z'$s
with the average \lcdm\ halo mass aggregations histories.  Using this connection, it is possible
to infer the \ms\ growth histories from the SHMRs. From these histories
we can find an active to passive transition mass at each $z$, i.e., the epoch when the stellar mass growth 
was almost halted. Since here we use the same SHMRs than in \citet{Firmani+2010}, 
the average transition mass is roughly the same as in that paper.  
Galaxies above the black solid line in Fig. \ref{Mtrans-z} 
are on average passive while those below are mostly active 
(in the sense of stellar mass growth, which can be associated mainly to the star formation activity). 

From Fig. \ref{Mtrans-z} we learn that at redshifts higher than $z\sim 0.5$, the galaxies in the
process of  transforming into bulge dominated are mostly still active star forming (blue) galaxies. 
For lower redshifts, the morphological transitions happen mostly already in the passive regime
of these (red) galaxies, likely through dry mergers.  
This picture may be supported by the observational results of 
\citet[][see also \citealt{Pozzetti+2010}]{Moresco+2013}
at $z < 1$. By using different definitions for early-type galaxies, they 
suggest that these galaxies, at masses $\ms < 10^{11}$ \msun,    
first experienced a transition in color from blue to red, and then in morphology. 
On the other hand, at $z > 1$, \citet{Talia+2013} 
found that only $\sim 33\%$ of all their morphological ellipticals are red 
and passive galaxies, while the rest of these ellipticals are star-forming galaxies. 
This suggests that morphological transformations at $z > 1$ are occurring 
before the transitions in star formation activity (or color), 
as indicated in Fig. \ref{Mtrans-z}.

\section{Comparison with observations}
\label{observations}

Before comparing our results in more detail with direct observations, it should be emphasized that 
the bulge/disc decomposition of observed galaxies is a very difficult task
\citep[see e.g.][]{Graham2001,MacArthur+2003,Allen+2006,Laurikainen+2007,
Fisher+2008,Tasca+2011,Simard-11}. On the other hand, since the \bt\ ratio and morphology of 
a galaxy depend on its luminosity (mass), determinations of the \bt\ distribution or the morphological mix are
strongly constrained by the completeness of the studied sample. Due to these difficulties, 
there are only a few studies of bulge/disc decomposition applied to local volume-limited samples
that can be used to obtain fair statistics on the \bt\ ratio as a function of \ms. The situation
is much worse at higher redshifts (see below).
Another possible issue when comparing with observations is that our results are only 
shown for central galaxies, whereas observational results can also include satellite galaxies. 
However, the total number fraction of satellites is relatively low ($20-25\%$) at $z \sim 0.1$ 
\citep[][]{Yang+2007}.  This fraction decreases drastically to higher redshifts
\citep{Knobel+2012}, so that the contamination of satellites at high $z$ is not expected 
to be relevant, specially for massive galaxies.
In the following, we attempt to compare our results with the
few observational studies on \bt\ statistics and evolution as a function of \ms; at high redshifts,
most works report indicators of morphology rather than \bt\ ratios, in such a way that we
have to roughly associate these indicators with corresponding \bt\ values.

\subsection{Local galaxies}

In Fig. \ref{BTvsMs-local}, where we plotted the \bt\ ratio vs \ms\ at $z\sim 0$ for our semi-empirical 
galaxies, we also reproduce observational results for two local volume-limited 
samples of galaxies: (i) $\sim 1000$ galaxies from the
SDSS with $\ms\geq10^{10}$ \msun\ \citep[][circles connected by the red line; only their central
galaxies were used, see Paper I for details]{Gadotti2009}, and (ii) 99 galaxies with $\ms\geq 10^9$ \msun\ 
in the local 11 Mpc volume \citep[taken from][squares connected by the blue line]{Fisher+2011}. Recent
determinations of the \bt\ ratio for a sample of $\sim 1100$ group galaxies ($z\sim0.05$, not from 
a volume-limited sample) presented in \citet{Carollo+2012} and \citet{Cibinel+2012} are 
also shown (\bt\ ratio in the $I$-band as black circles with error bars).
Overall our results follow the same trend than the observational inferences 
 \citep[see also the SDSS results by][]{Skibba+2012}, which actually 
have a large scatter and differ among them. It seems that for $\ms\gtrsim 10^{10}$ \msun, our \bt\ ratios 
are lower on average than observations. A more quantitative comparison is using the
\bt\ distribution (histogram) 
for volume-limited samples above a given \ms; as shown in Paper I
our results agree well with the few available observational samples, and even the fractions of classical 
and pseudo bulges are roughly reproduced. See also Section \ref{slowSHMR}.

In a recent paper, using the cross-match of the SDSS and RC3 catalogs given in \citet{Wilman+2012}, 
\citet{Wilman+2013} estimate the local fractions of elliptical galaxies as a function of  \ms. 
For $\ms> 3\times 10^{10}$ \msun, the overall fraction is $0.08\pm 0.01$; this fraction raises to 
$\approx 0.4$ for galaxies above $6\times 10^{11}$ \msun.
As these authors suggest,  elliptical galaxies can be associated to those with $\bt>0.7$. The fraction 
of our local galaxies more massive than $3\times 10^{10}$ \msun\ with $\bt>0.7$ is $0.12\pm 0.05$ and it
also increases if the mass threshold is increased.

\begin{figure}
\centering
\includegraphics[height=6.7cm,width=8.5cm,trim=1.2cm 0.5cm 0.5cm -0.5cm, clip=true]{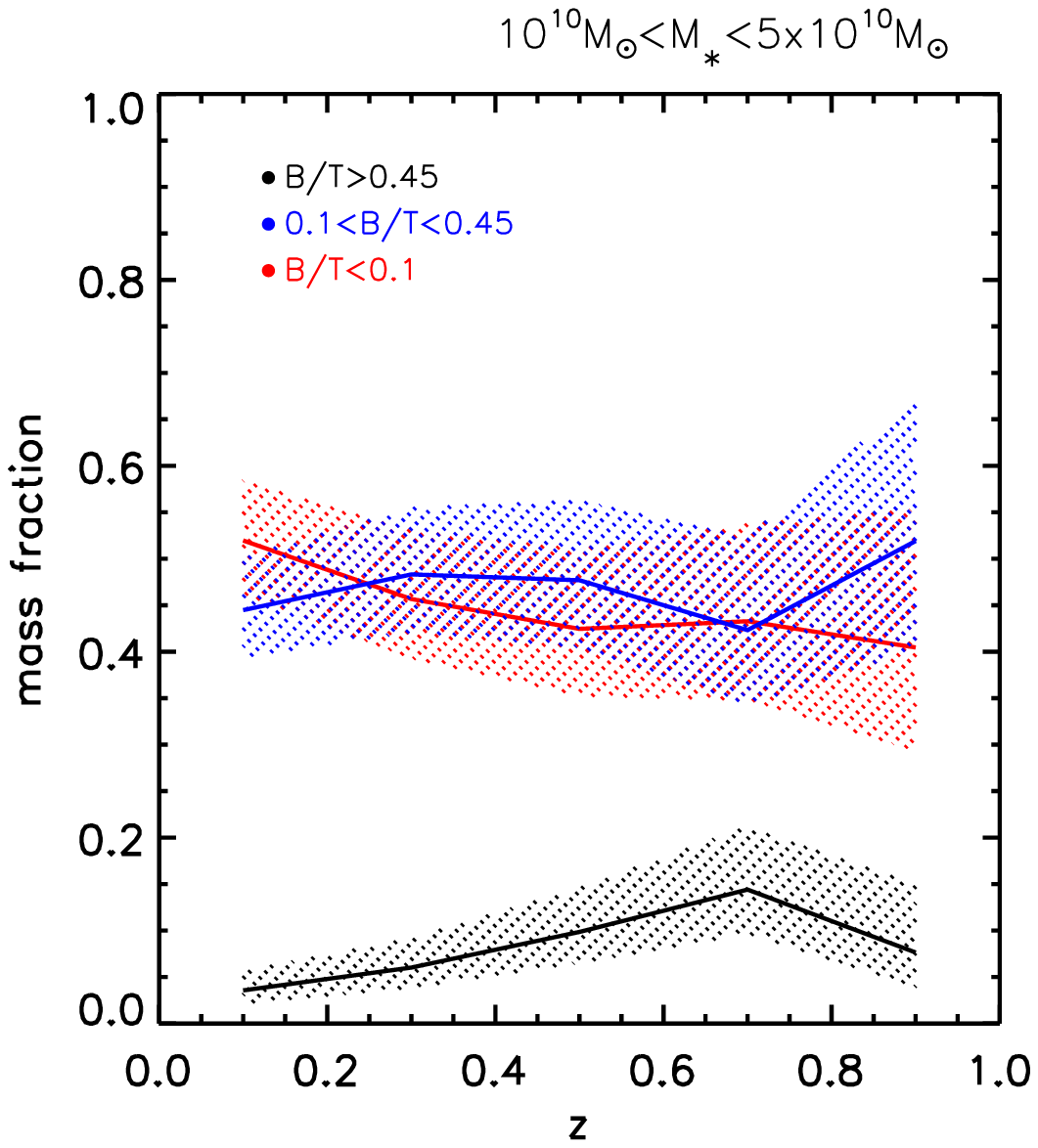}
\includegraphics[height=6.7cm,width=8.5cm,trim=1.2cm 0.5cm 0.5cm -0.5cm, clip=true]{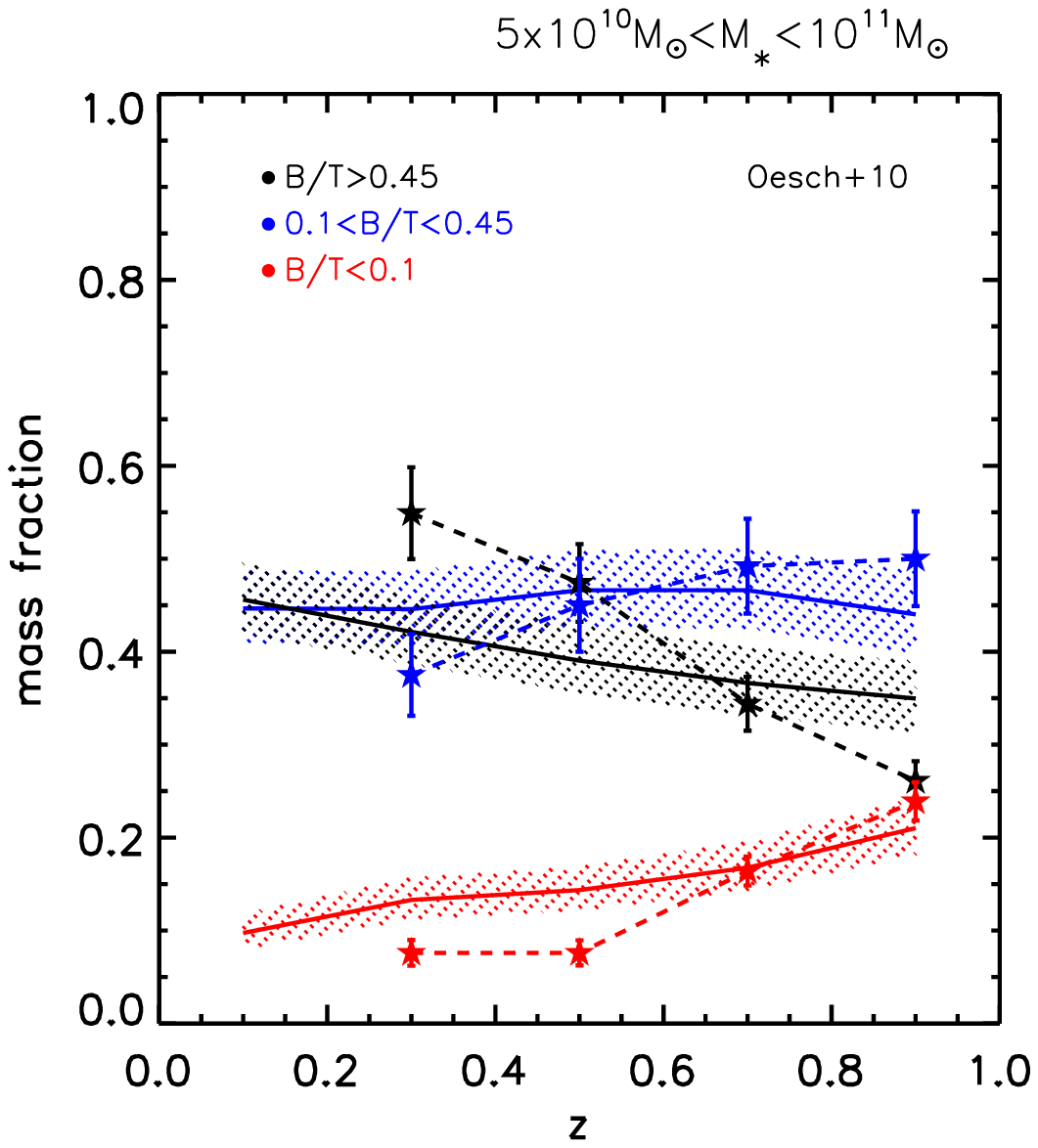}
\includegraphics[height=6.7cm,width=8.5cm,trim=1.2cm 0.5cm 0.5cm -0.5cm, clip=true]{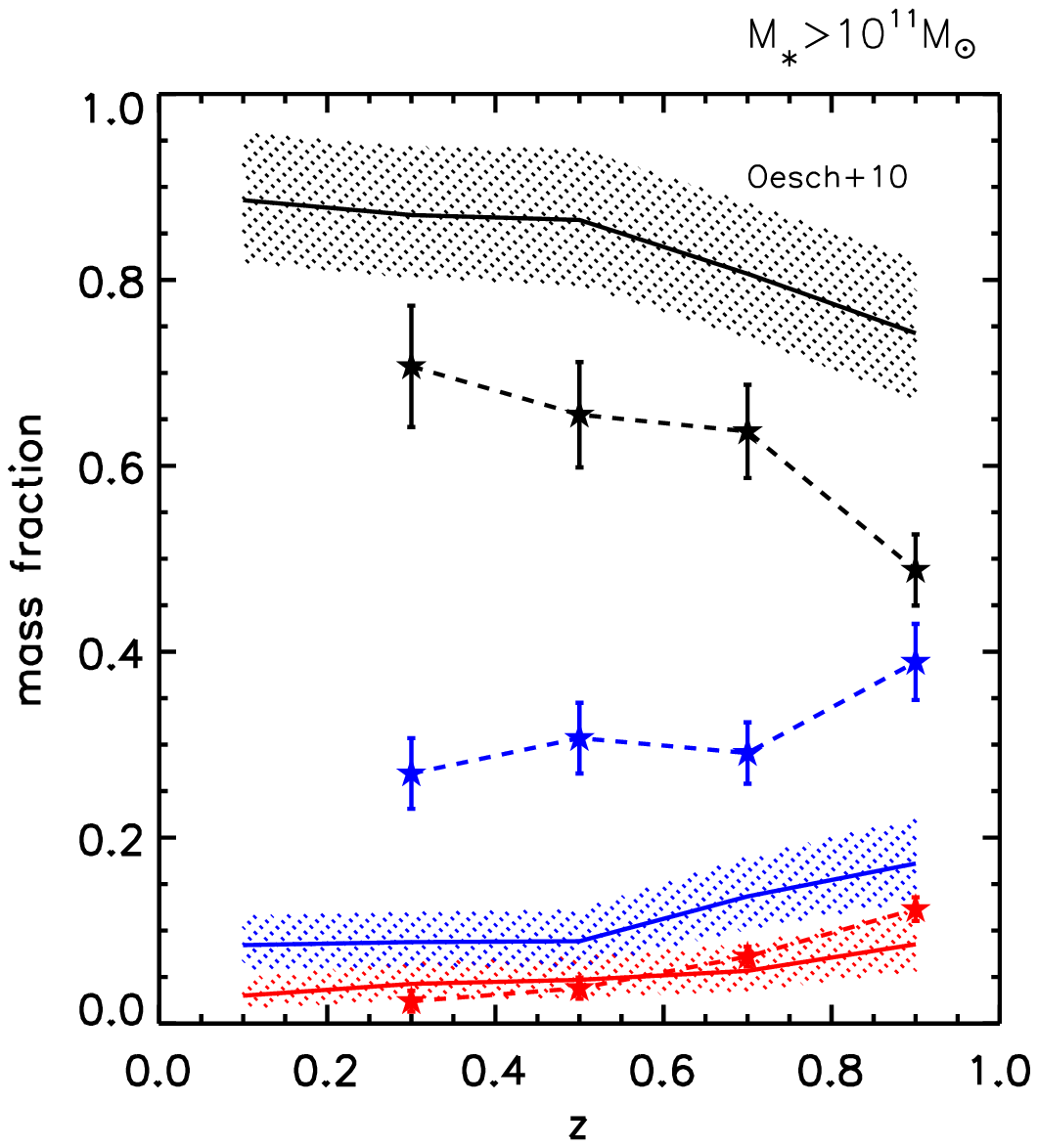}
\caption{\textit{Mass} fractions of galaxies according to their \bt\ values (indicated by a color
code in the legends) as a function of $z$ for the semi-empirical galaxies.
The dot-shaded areas show the Poisson errors.
The stars with error bars connected by short-dashed lines correspond to 
observations from $z$COSMOS \citep[][see text for the equivalences between the \bt\ ratios and
the morphological classes given by these authors]{Oesch+2010}.  The top, medium and bottom panels
are for different mass bins, as indicated above each panel. For the lowest mass bin, there
are no observational data. 
}
\label{Oesch}
\end{figure}

\subsection{High-redshift galaxies}
\label{high-z}

It is only in the last years that a few observational works have appeared reporting reliable morphologies, and even 
bulge/disc decompositions, for relatively large samples of massive galaxies at high redshifts by using 
high resolution images, mainly obtained with the Hubble Space Telescope. 
Unfortunately, 
the indicators used to define the morphologies are different 
among different works. Therefore, in order to compare the observational results with the predicted 
evolution of the \bt\ ratio, a rough equivalence for these different morphological 
indicators with the \bt\ ratio should be established.

Based on $\sim$8600 galaxies from the $z$Cosmic Evolution Survey \citep[$z$COSMOS;][]{Scoville+2007}, 
and using the Zurich Estimator of Structural Types (ZEST),
\citet[][]{Oesch+2010} studied the evolution of the morphological mix of galaxies at stellar masses
$>5\times 10^{10}$ \msun\ from $z=0.2$ to $z=1$.  ZEST is based on structural parameters 
such as ellipticity, concentration, asymmetry, the second-order moment of the light distribution, M$_{20}$, 
and the Gini coefficient \citep[][]{Scarlata+2007a}. \citet[][]{Oesch+2010}  grouped galaxies into five 
morphological classes. Here we re-group these galaxies into three broader classes and assign to these classes
a range of \bt\ ratios in order to compare them with our semi-empirical galaxies: elliptical (E) and
bulge-dominated (B) galaxies are assigned to a first group with $\bt>0.45$;  spiral galaxies with
intermediate bulge properties (S) are assigned to a second group with $0.1<\bt\leq 0.45$;
disc-dominated (D) and irregular (I) galaxies are assigned to a third group with $\bt\leq 0.1$.

\citet[][]{Oesch+2010} report the evolution of the \textit{mass} fractions corresponding to different morphological 
classes for two mass bins. The medium and bottom panels of Fig. \ref{Oesch} reproduce the \citet[][]{Oesch+2010} 
results re-grouped into the three aforementioned groups; black, blue, and red stars with error bars (connected by
dashed lines), respectively. The corresponding mass fractions
from the semi-empirical model are shown with black, blue and red solid lines, respectively.
The dotted-dashed regions are Poissonian errors in the number counts.
In the top panel we show the model predictions for smaller galaxies. The general trends of the mass fractions 
with $z$ and \ms\ are similar between the predictions and the observational results of \citet[][]{Oesch+2010},
as can be seen in the medium and bottom panels. 
The morphological mix from observations changes moderately from $z\sim 1$ to $z=0.2$:  
the fraction of bulge-dominated galaxies increases towards lower $z$, while the fraction of other classes 
decreases. The semi-empirical results show, overall, less relative evolution of the morphological mix 
than observations, and a stronger dependence of the mass fractions with mass. 
The main difference between predictions 
and observations is that the mass fraction of massive, $\ms>10^{11}$ \msun, bulge-dominated ($\bt>0.45$) 
galaxies is higher by 2-3~$\sigma$ in the model than in observations (bottom panel).

Using the $z$COSMOS data and ZEST, \citet{Kovac+2010} determined the number fraction
of early-type galaxies (E+B types), $f_{\rm early}$, as a function of \ms\ at different redshifts 
and in different environments.  
As above, we assume that E+B types correspond to $\bt>0.45$ and plot the 
\citet{Kovac+2010} results for field galaxies in the panels of Fig. \ref{BT-histograms}
with the closest $z$ bins
to those reported by these authors ($0.2<z<0.4$, $0.4<z<0.6$, and $0.6<z<0.8$). 
The trends with mass at different redshifts are the same for observed and semi-empirical galaxies, 
although the former have higher fractions of early-type galaxies than the latter, showing that
the fractions of massive bulge-dominated galaxies in the $z$COSMOS sample are somewhat
higher than in our mock catalog.

\begin{figure}
\centering
\includegraphics[height=7.5cm,width=8.5cm,trim=1.2cm 0.5cm 0.5cm -0.5cm, clip=true]{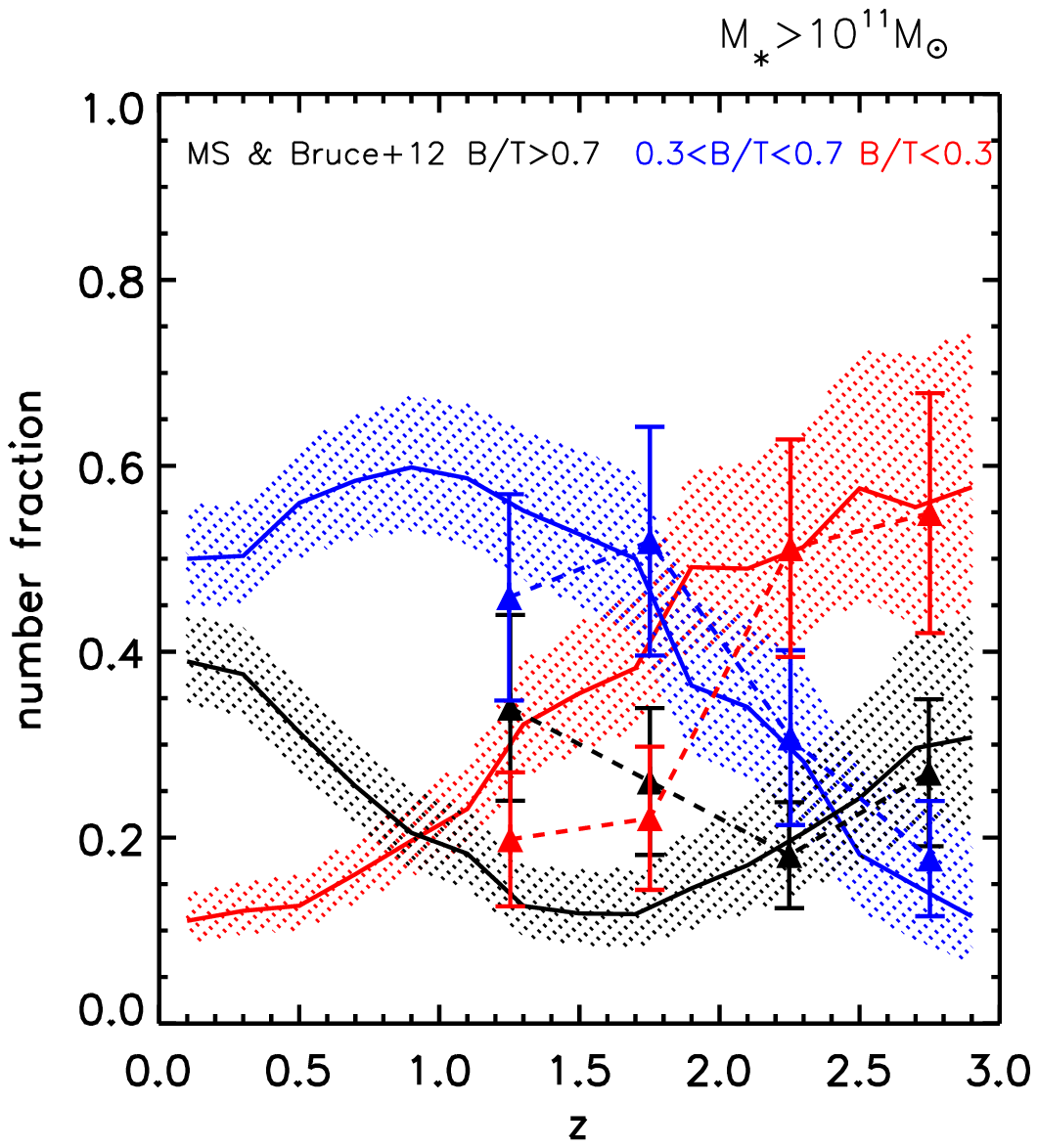}
\caption{Number fraction of the semi-empirical galaxies more massive than $10^{11}$ \msun\ according 
to their \bt\ values (indicated by the color code in the legend) as a function of $z$ (solid lines). 
The dot-shaded areas bracket the Poisson errors of the number counts. 
The triangles with error bars connected by short-dashed lines correspond to \bt\ ratios 
measured from galaxies observed from $z\sim 1$ to  $z\sim 3$ by \citet{Bruce+2012}. }
\label{Bruce}
\end{figure}

\begin{figure}
\centering
\includegraphics[height=7.5cm,width=8.cm,trim=1.2cm 0.5cm 0.5cm -0.5cm, clip=true]{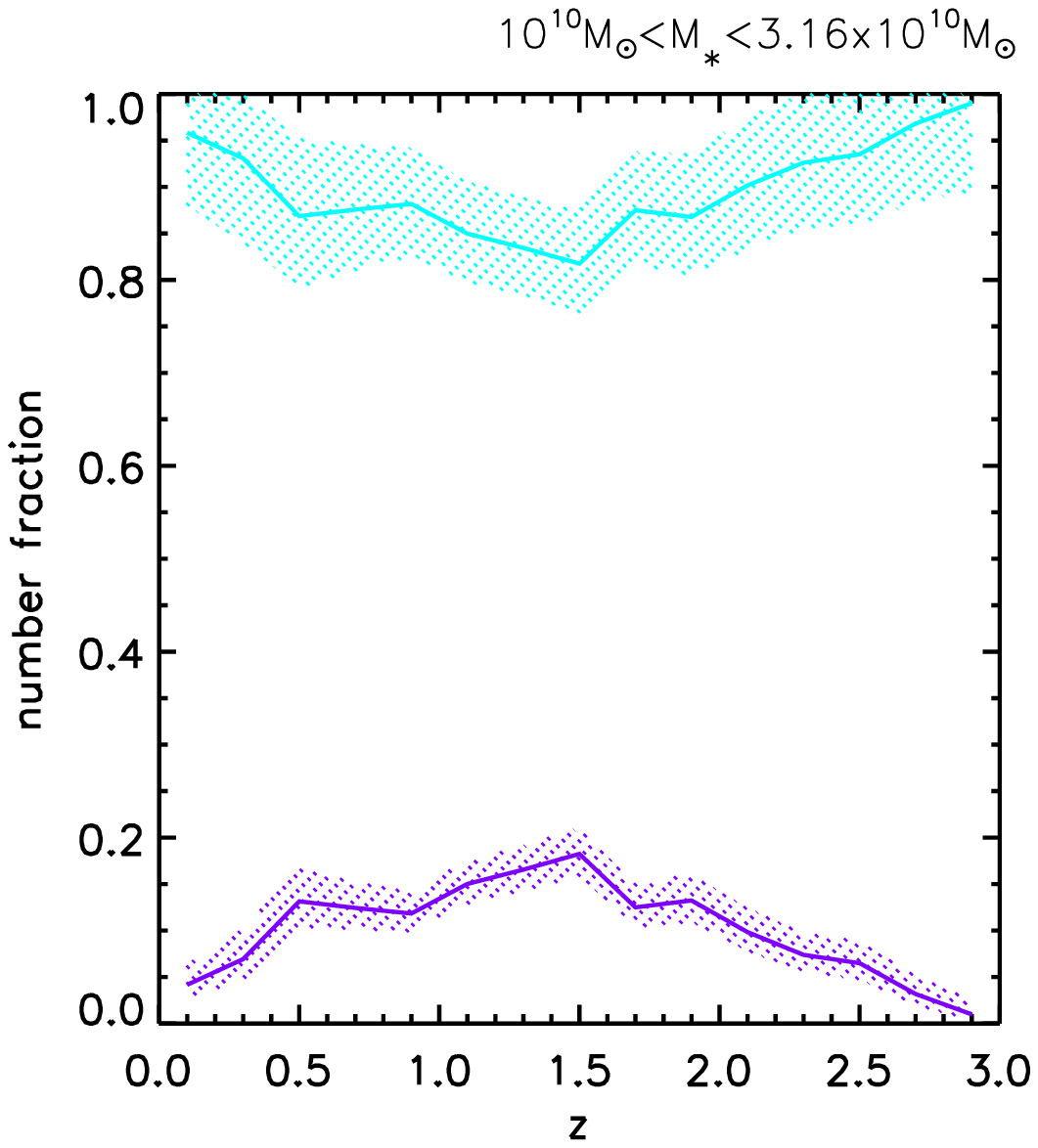}
\includegraphics[height=7.5cm,width=8.cm,trim=1.2cm 0.5cm 0.5cm -0.5cm, clip=true]{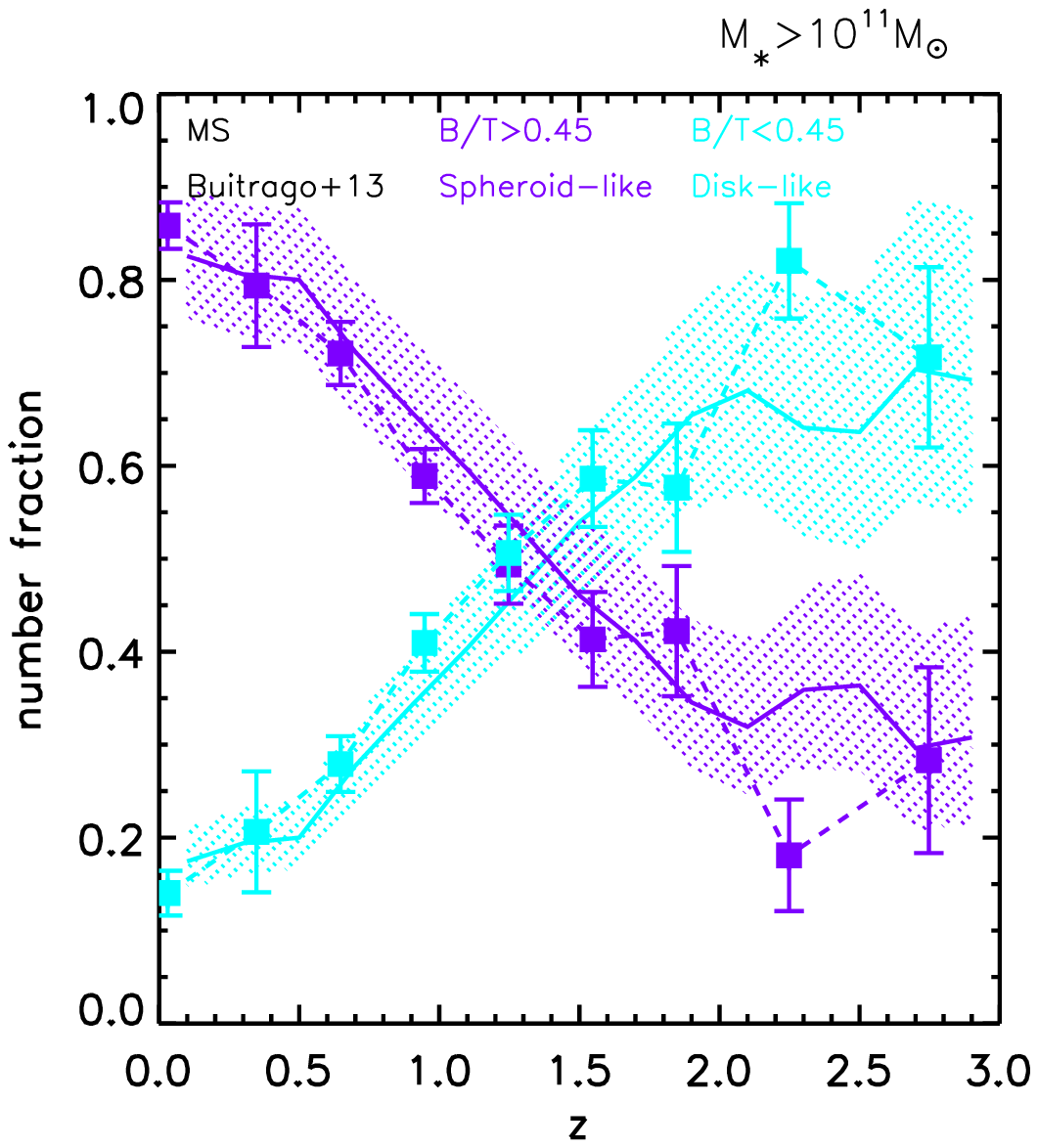}
\caption{Number fraction of bulge- and disc-dominated semi-empirical galaxies (purple and cyan colors, respectively) 
as a function of $z$, from $z\sim 0$ to $z\sim 3$. The upper (lower) panel is for galaxies 
of masses $10^{10} - 3.16\times 10^{10}$ \msun\ ($> 10^{11}$ \msun). 
The dot-shaded areas bracket the Poisson errors of the number counts. 
The squares with error bars connected by short-dashed lines correspond to observations 
by \citet[][]{Buitrago+2013}. }
\label{Buitrago}
\end{figure}

 At $z>1$, morphological studies have only been done for massive galaxies.
 In a recent work, \citet{Bruce+2012} fitted bulge/disc models to the $H_{160}$-band images of $\sim$200 
 massive galaxies (\ms\ $>$ 10$^{11}\msun$) at 1 $< z <$ 3 from the CANDELS-UDS field. From their
 formally-acceptable fits, the \bt\ ratios of $>90\%$ of the sample was obtained. 
 In Fig. \ref{Bruce} we plot the evolution of the \textit{number} fractions corresponding to
 three \bt\ bins as reported in \citet{Bruce+2012}: $\bt\leq 0.3$ (red triangles connected
 by the red dashed line),  $0.3<\bt\leq 0.7$ (blue triangles connected by the blue dashed line), 
 and $\bt>0.7$ (black triangles connected by the black dashed line). The error bars show the $1\sigma$ scatter. 
 The corresponding results from our semi-empirical model (down to $z\sim 0$)
 are shown with the solid red, blue, and black lines, respectively; the dot-shaded bracket the Poissonian errors 
of the number counts.
 
 According to Fig. \ref{Bruce}, the trends in the morphological mix evolution 
of the semi-empirical and observed massive galaxies are quite similar.  One can say 
that \textit{the redshift range $2<z<3$ is the era of massive discs}. In this redshift range,
a substantial fraction of both our semi-empirical (see Fig. \ref{BT-histograms}) and observational 
\citep[see][]{Bruce+2012} massive galaxies 
are almost pure discs, $\bt<0.1$. In the range $1<z<2$, the fraction of massive pure discs 
systems falls dramatically in favor of disc+bulge systems.  At $z\sim 1$, while bulge-dominated 
systems are on the rise, galaxies comparable to present-day giant ellipticals are 
a minority. Note that from $z\sim 2$ to $z\sim 1$, the fraction of systems 
with $\bt>0.7$ ($\bt<0.3$) increases (decreases) more in the observational sample than in our case. 
However, at $z<1$ the fraction of semi-empirical massive galaxies with $\bt>0.7$ rises 
strongly.  

\citet{Buitrago+2013} reported the morphological mix evolution from $z=3$ to $z\sim 0$ for 
massive galaxies, $\ms>10^{11}$ \msun, using a statistically representative sample of nearly 
$1000$ galaxies from the SDSS, the Palomar Observatory Wide-field InfraRed/DEEP2, 
and the GOODS NICMOS surveys. These authors applied a qualitative visual 
morphological classification in addition to a quantitative estimate based on the S\'ersic index $n$.
The latter parameter is well correlated with \bt\ in the sense that higher 
$n$ values correspond to higher \bt\ ratios \citep{Bruce+2012}. In the lower panel of
Fig. \ref{Buitrago} we reproduce the \textit{number} fraction evolution of the \citet{Buitrago+2013} 
sample divided in two groups, disc-dominated galaxies ($n<2.5$; cyan squares with error bars) 
and bulge-dominated galaxies ($n>2.5$; purple squares with error bars). We identify the former 
objects as those with $\bt<0.45$ and the latter as those with $\bt\geq 0.45$ and plot the
corresponding fractions as a function of $z$ for our semi-empirical sample of galaxies
(cyan and purple lines, respectively). The corresponding dot-shaded areas show the
Poisson errors of the number counts.

The agreement in the morphological evolution of massive galaxies between our 
semi-empirical model and the observations reported in \citet{Buitrago+2013} is
remarkable. The fraction of bulge-dominated galaxies among the massive galaxy population 
has increased from $20-30\%$ at $z=3$ to $\sim$80\% at $z=0$. 
Bulge-dominated galaxies have been the predominant morphological
class for massive galaxies only since $z\sim 1$ (see also Fig. \ref{BT-histograms}). 
From the visual morphological classification, \citet{Buitrago+2013} find that a 
fraction of their sample are merging/peculiar galaxies; this fraction
is very low at low redshifts but it increases from $\sim 10\%$ at $z=1$ to $\sim 35\%$ at $z\sim 3$. 
Most of these galaxies seem to correspond to those with $n<2.5$. In the case
of the massive semi-empirical galaxies, we find that the fraction of those
suffering a major merger is similar to the one reported in \citet[][see Paper I and elsewhere for 
detailed comparisons with observations]{Buitrago+2013}.

The upper panel of Fig. \ref{Buitrago} is analogous to the lower one but
for the semi-empirical galaxies in the $10^{10}<\ms/\msun< 3\times 10^{10}$ mass bin. 
The fraction of bulge-dominated galaxies increases from virtually $0\%$ at $z=3$
to $20\%$ at $z\sim 1.5$ and then again decreases, reaching $\sim 5\%$ at $z=0$.  
As has already been seen in Figs. \ref{BT-histograms} and \ref{BTevolution}, a fraction of the
low-mass galaxies may have attained a significant \bt\ ratio by $z\sim 1-1.5$ but afterwards,
the major merger rates at these scales are negligible in such a way that the ulterior 
(significant) \ms\ growth happens only for the disc. 

In a recent paper, \citet{Mortlock+2013} extended the morphological classification of 
galaxies to lower masses ($\ms >10^{10}$ \msun) at $z > 1$ by using the S\'ersic index $n$ 
reported by \citet{vanderWel+2012} for $\sim 1100$ galaxies from the CANDELS/UDS field.
The number fractions of bulge-dominated ($n>2.5$) and disc-dominated ($n<2.5$) galaxies with masses 
$10^{10}<\ms/\msun< 3\times 10^{10}$ remain roughly constant from $z\sim 3$ to $z\sim 1$. 
These fractions are approximately 70\% and $25\%$ 
for the former and latter galaxy types, respectively. A small fraction
of galaxies ($\sim 5\%$) have undetermined $n$ index. These results are qualitatively similar to those
shown in the upper panel of Fig. \ref{Buitrago}, although the fractions of bulge-dominated (disc-dominated) 
are higher (lower) for the observations than for the models. 
For masses larger than $3\times 10^{10}$ \msun, \citet{Mortlock+2013} find that the disc-dominated 
galaxies are more abundant than the bulge-dominated ones down to $z\sim 1.5-2$. At
lower redshifts, the bulge-dominated galaxies start to be more abundant, in agreement 
with \citet{Bruce+2012} and \citet{Buitrago+2013}, and therefore with our results.

\section{Discussion}
\label{discussion}

\begin{figure}
\includegraphics[height=7.2cm,width=8.4cm,trim=-0.2cm 0.cm 0.cm -0.cm, clip=true]{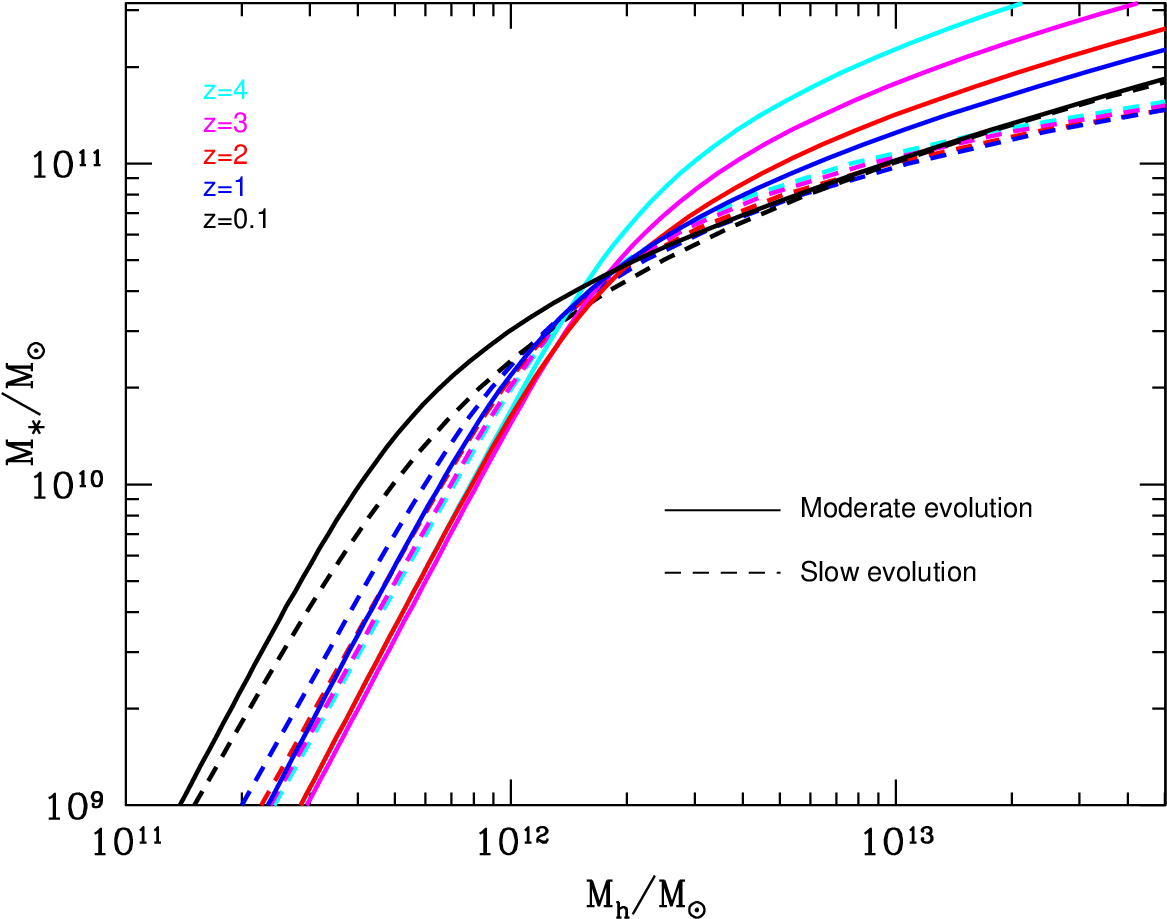}

\caption{The SHMR relation at four redshifts as indicated in the legends. Solid lines
are for the ``moderately evolving'' SHMR \citep{Firmani+2010} and dashed lines are for
the ``slowly evolving'' SHMR, resembling closely the results by \citet{Behroozi+2013-L}. }
\label{SHMR}
\end{figure}

\subsection{The case of a ``slowly evolving'' SHMR}
\label{slowSHMR}

An important ingredient in our scheme is the empirically constrained SHMR that we use at each redshift to 
assign stellar masses to the distinct haloes from the Millennium simulations (Section \ref{model}). 
 We have used the SHMR parametrization given and constrained by \citet{Behroozi+2010} and slightly modified
by \citet{Firmani+2010} in order to make it continuos from $z=0$ to $z=4$ (see Appendix A).

Several new constraints on the SHMR at different redshifts have appeared recently 
(e.g. \citealp{Yang+2012}, \citealp{Leauthaud+2012}, \citealp{Wake+2011}, \citealp{Moster+2013},
\citealp{Behroozi+2013-L}, \citealp{Behroozi+2013-main}, \citealp{Wang+2013}). Some of them 
present a stronger evolution with $z$ than the SHMR used here, while others evolve less.
According to \citet{Behroozi+2013-L}, the SHMR changes little from $z=0$ to $z\sim 4$. In order to explore
the effects of the adopted SHMR evolution on the demographics and evolution of the \bt\ ratio, we 
obtained new results changing the evolution of the \citet{Firmani+2010} SHMR parameters in such a way that
the \citet{Behroozi+2013-L} SHMRs are closely reproduced from $z=0.1$ to $z\sim 4$ in the halo mass
range $10^{11}<\mh/\msun < 10^{13}$ (dashed lines in Fig. \ref{SHMR}; see Appendix A for the parameter
values of this SHMR). 

\begin{figure}
\includegraphics[height=7.2cm,width=8.4cm,trim=-0.2cm 0.cm 0.cm -0.cm, clip=true]{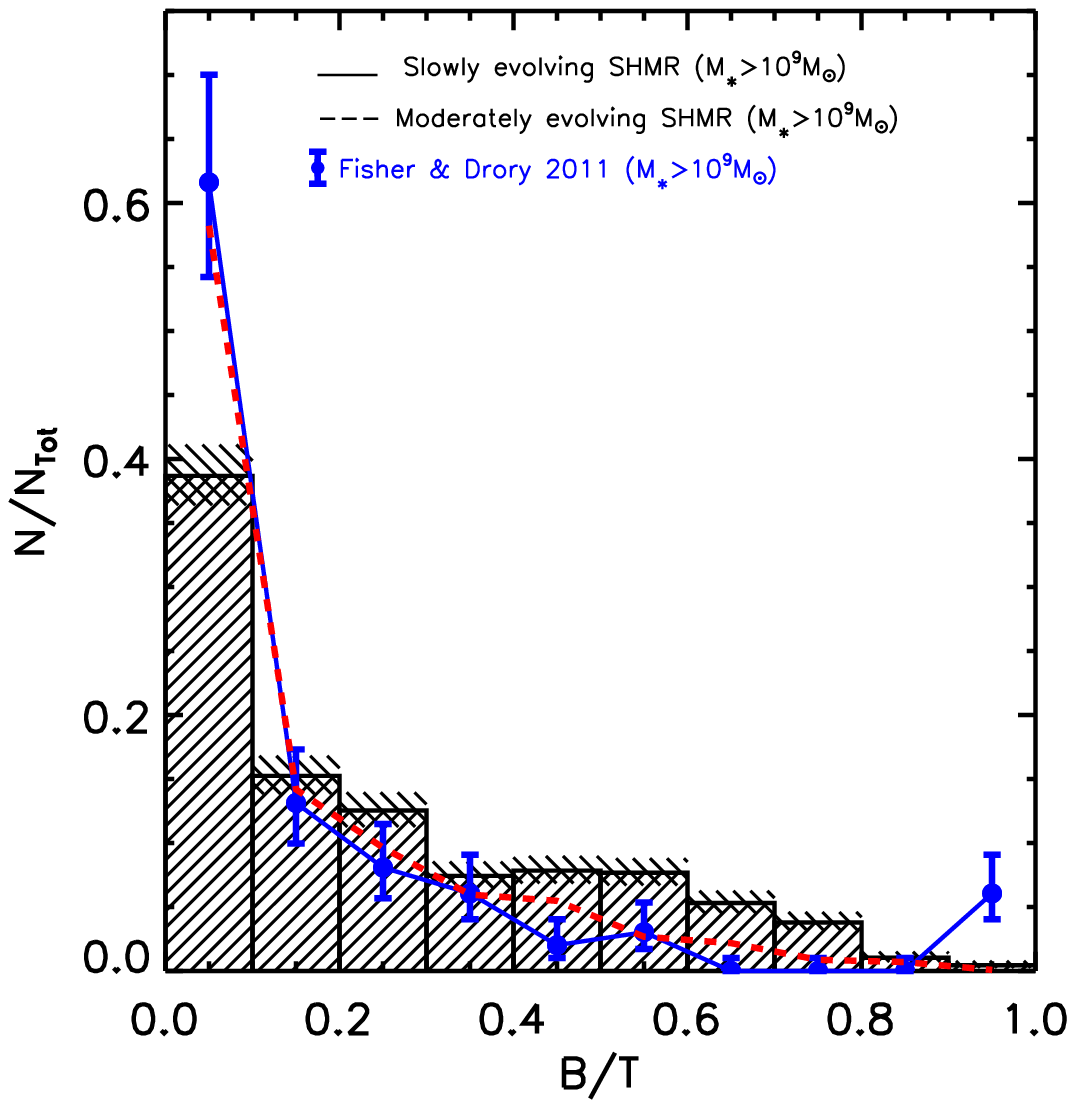}

\caption{$B/T$ distribution for semi-empirical galaxies with $M_{\ast}\geq10^{9}$M$_{\odot}$ 
using two different SHMRs for the models:  ``slowly evolving'' (black histograms) and ``moderately 
evolving'' (red dashed line; fiducial model). The corresponding distribution from the the observational 
sample from \citet{Fisher+2011} is shown with blue symbols. Errors in the number counts are 
Poissonian and are marked with bars for the observational data and with shaded regions for the 
``slowly evolving'' SHMR model (the amplitude of the errors is similar for the fiducial case).}
\label{BT_histogram_SHMR}
\end{figure}

\begin{figure}
\centering
\includegraphics[height=10cm,width=8.8cm,trim=0.2cm 0.2cm 0.2cm -0.2cm, clip=true]{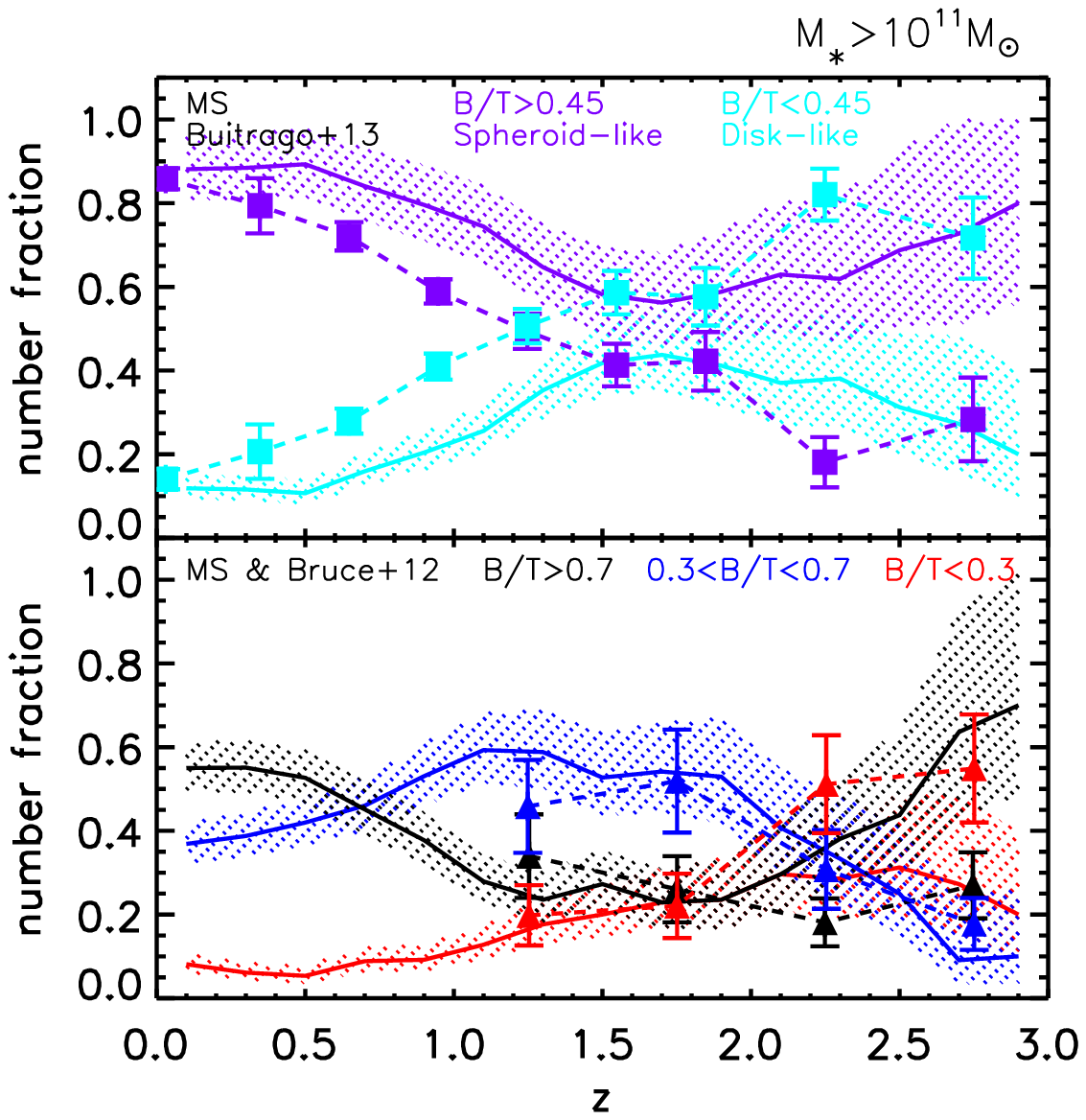}
\caption{As in Figs. \ref{Bruce} and  \ref{Buitrago}, but for our modelling obtained 
with the ``slowly evolving'' SHMR instead of the ``moderately evolving'' SHMR (fiducial case).
}
\label{model2}
\end{figure}

In general, for the ``slowly evolving'' SHMR, the \bt\ ratios are higher (specially at low masses) and the 
bulges assemble earlier than for the ``moderately evolving'' SHMR (fiducial case, Sections \ref{results} 
and \ref{observations}). Fig. \ref{BT_histogram_SHMR} shows the \bt\ distribution for the mock galaxies 
with $M_{\ast}\geq10^{9}$M$_{\odot}$ in these two cases. Observations from the \citet{Fisher+2011} sample
are also shown with blue symbols with error bars. It is clear that the ``slowly evolving'' case is at odds 
with this observational sample, producing too few bulgeless galaxies, contrary to the ``moderately evolving'' 
case that is in remarkable agreement with the \citet{Fisher+2011} observations (Paper I). 
However, at intermediate masses ($10^{10}<\ms/\msun < 10^{11}$), the ``slowly evolving'' case is
in slightly better agreement than the fiducial case with the observational results from \citet{Gadotti2009} 
and \citet{Cibinel+2012}. 
For massive galaxies, $\ms>10^{11}$ \msun, the ``slowly evolving'' SHMR produces too many bulge-dominated 
galaxies compared to the results of \citet{Buitrago+2013} and \citet{Bruce+2012}, specially from $z=3$
to $z\sim 1.5$ (see Fig. \ref{model2}). For this case, the period $2<z<3$ is not the era of massive discs 
as observations suggest. The mass fractions of galaxies with $\bt>$ 0.45 ($0.1<\bt\leq 0.45$) from $z=1$ to 
$z=0.2$ are also significantly higher (lower) than the observational results of \citet[][]{Oesch+2010}; 
in particular, at $z\sim 1$, the fraction of $\bt>$ 0.45 galaxies in the mass range 
$5\times 10^{10}< \ms/\msun <10^{11}$ is two times higher than in \citet{Oesch+2010}. 

Let us understand why the ``slowly evolving'' SHMR produces results different to the
``moderately evolving'' SHMR used in Section 3. 
For a given \ms, at the high-mass end of the SHMR and at high redshifts, the halo mass is significantly larger 
for the former than for the latter (see Fig. \ref{SHMR}). 
This implies a larger fraction of high \bt\ galaxies at high $z$ because of two effects: (i) the merger rate
is higher for more massive haloes (see Paper I), and (ii) for a given $z$, the peak height of the density fluctuations corresponding to more massive haloes is higher, thus, there is a larger fraction of haloes with $\nu>2$ at $z_{\rm seed}$. 
Recall that we impose as our initial condition that if $\nu>2$ at $z_{\rm seed}$ for a given halo, then its central 
galaxy is born with \bt=0.9 (see subsection \ref{initial-conditions}). 

In the low-mass side of the SHMR, $\mh\lesssim 10^{12}$ \msun,
the galaxies grow increasingly faster with time towards lower masses \citep[downsizing in sSFR;][]{Firmani+2010}. 
This behavior is more dramatic for the ``moderately evolving'' SHMR used in the previous sections 
than for the ``slowly evolving'' SHMR. Therefore, while the discs continue growing in the former case, 
making the \bt\ ratios smaller, in the latter case the low-mass galaxies grow less, keeping their relatively high
\bt\ ratios, acquired early during the active merging epochs. As a result, for low-mass galaxies formed in the
``slowly evolving'' SHMR case: (i) their \bt\ ratios are higher today and (ii) the morphological
mix changes much less since $z\sim 1$ than in the case of the ``moderately evolving'' SHMR 
used in previous sections as our fiducial case.

\subsection{Is the \lcdm-based bulge growth consistent with observations?}
\label{lcdm}

The semi-empirical model of bulge growth presented in Paper I and here is based on the
cosmological \lcdm\ scenario, specifically, it rests on the merger rates as a function of time
that galaxies suffer insider the growing CDM haloes, 
with the spheroidal component (bulges) assembling as the
result of merger-driven processes. Bulges acquire their stars from the merged secondaries, 
from the primary disc due to instabilities induced by the mergers (even those that almost do 
not contribute with stars but perturb the disc with their dynamical masses), as well as through 
stars formed in situ from the gas that is funneled to the center during mergers.  These 
different channels may give rise to classical- and pseudo-like bulges residing 
in the same galaxy, \textit{i.e., bulges can be actually composite,}  with the pseudo bulge 
component being the product of merger-driven instabilities.  By means of numerical
simulations of S0 galaxies, \citet{Eliche-Moral+2013} have shown that intermediate and major 
mergers indeed trigger significant internal secular evolution in the discs (difficult to isolate 
from the purely intrinsic disc instabilities), which is seemingly able to preserve the 
structural coupling of the bulge and the disc. 

It is important to mention that in our model the galaxies formed in rare high-$\sigma$ 
massive haloes at $z_{\rm seed}\sim 3.5$ are seeded as bulge dominated (see subsection \ref{initial-conditions}), 
resembling a monolithic scenario rather than the hierarchical one. Since the stellar masses of 
massive galaxies almost do not grow further according to the semi-empirical SHMRs, the assumed 
high \bt\ ratios for these galaxies remain in most of cases as such until the present day.

In Section \ref{observations}, we have compared the semi-empirical results corresponding to our
fiducial case with currently available direct observational studies of the morphological mix of galaxies at different redshifts.  
 \textit{All the observed general trends of the fractions of galaxies with a given \bt\ ratio as a function of 
\ms\ and $z$ are in good agreement with our results.} At this level, the $\lcdm$-based semi-empirical 
approach we have used here to estimate the growth of bulges does not seem
to face critical issues. At a quantitative level, 
our results are in most cases consistent with these observations, within the large systematic and statistical uncertainties.

However, we have found also a few quantitative discrepancies that should be discussed.
Before that, it is worth noting that (i) observational studies at high 
redshifts typically report different morphological classes 
rather than \bt\ ratios, and (ii) a criterion of morphological classification in the local universe 
is not always useful at higher redshifts. For example, \citet{Talia+2013} found that the 
parameters of asymmetry and M$_{20}$ are not effective in distinguishing morphologies 
at $z > 1$, in contrast to what is observed at $z \sim$ 0.  A similar result is found by \citet{Mortlock+2013}, 
thus suggesting that high $z$ galaxies are structurally different from their counterparts at 
low $z$. 
In general, disc galaxies are misclassified as spheroids due to the lower resolution 
of the images at high redshift, which removes the signatures of a disc structure. Therefore, it
is possible that the observational fractions of bulge-dominated galaxies are overestimated at high redshifts.
 
The most noticeably quantitative difference among the semi-empirical galaxies and observations 
is in the \textit{mass} fractions of massive galaxies, $\ms > 10^{11}$ \msun, between $0.2 < z < 1$ 
(bottom panel of Fig. \ref{Oesch}, observations from \citealt{Oesch+2010}). Although qualitatively for both, 
models and observations, the mass fraction of massive objects is dominated by galaxies with 
an important bulge component, quantitatively, the mass fraction of model galaxies with $\bt > 0.45$ 
($0.1 < \bt < 0.45$) is higher (lower) than that of the corresponding observed morphological class by 
2-3$\sigma$. This discrepancy diminishes at lower redshifts. 
Note that the \textit{mass} fraction takes into account the number fraction and the number density of galaxies.
Since massive galaxies are rare, cosmic variance has probably an important role in
their number densities. 
For the same zCOSMOS sample, the \textit{number} fractions of bulge-dominated 
massive galaxies as reported in \citet{Kovac+2010} are actually larger than in our case 
(see Fig. \ref{BT-histograms}).

At intermediate masses, $10^{10} < \ms/\msun < 3.2\times10^{11}$, our fiducial 
model predictions show that the number fraction of disc-dominated galaxies completely dominates at all redshifts
(upper panel of Fig. \ref{Buitrago}). This is in qualitative agreement with the observational results of \citet{Mortlock+2013} 
reported at 1 $< z <$ 3. However, the model predicts a fraction greater than 0.8 for galaxies with \bt$<0.45$, 
while the fraction of observed galaxies with S\'ersic index $n <$ 2.5 (disc-dominated systems) is not 
higher than 0.8 at any $z$.  Finally, at low redshifts, the model seems to predict slightly lower 
\bt\ ratios of galaxies of intermediate masses than observations (see Fig. \ref{BTvsMs-local} and the 
middle panel of Fig. \ref{Oesch}); there is also a possible slight excess of pseudo-bulges 
over classical bulges as compared with the local observations (Paper I).  

Interesting enough, the \lcdm-based semi-empirical fiducial model, instead of predicting an excess of high-\bt\ 
classical bulge galaxies at intermediate masses with respect to observations, it seems to predict a slight 
deficit of them. Recently, alternative mechanisms of bulge formation have been proposed. For example, 
the fragmentation of the gas-rich disc into clumps that migrate towards the centre can form a large 
spheroid in intermediate-mass galaxies at $z\sim 1-2$ \citep{Dekel+2009, Perez+2013}.  According to 
the latter authors, this spheroid has the features of a classical bulge. This extra mechanism could perfectly 
fit in our scheme, increasing the fraction of higher \bt\ intermediate-mass galaxies from $z\sim 3$ to $z\sim 1$. 
If the bulges thus formed of some of these galaxies do not increase much more down to $z=0$, then they 
will contribute to the classic-bulge dominated population today.Ê

\section{Conclusions}
\label{conclusions}

The mass aggregation and merger histories of a subsample  of present-day distinct haloes from the
Millennium Simulations are used to calculate the stellar mass growth and merger histories of galaxies.
Galaxies are seeded at the centre of the distinct haloes (and in subhaloes at the 
accretion time) by means of the stellar-to-halo and gas-to-stellar mass relations constrained by 
observations at different redshifts. The merger-driven bulge 
growth of these galaxies is calculated by using physically motivated recipes, which account for
three channels of bulge mass acquisition: stars from the merged secondary, stars 
transferred from the primary disc due to instabilities induced by the merger, and stars 
formed from gas funneled from both merging galaxies.  At intermediate masses, the first and second channels combine
in such a way that the bulges are actually \textit{composite} (Paper I). At small masses, the
second channel dominates by far, producing pseudo-like bulges, while at large masses, the 
first channel dominates, producing classical-like bulges. 

Our semi-empirical model offers a transparent way to map the \lcdm\ halo mass accretion and 
merger histories to the stellar mass growth of the galaxies and their bulges.  
In the following, we present the main results and conclusions obtained with this semi-empirical 
model using a SHMR that moderately changes with $z$.

\begin{itemize}

\item The morphological (\bt\ ratio) mix at different stellar masses remains qualitatively the same
since $z\sim1$, while for $z>1-1.5$, it changes towards a larger 
population of disc-dominated and bulgeless galaxies. In the $0<z<1$ period, the most
abundant galaxies are: bulgeless ($\bt\leq0.1$) at low masses, $\ms\lesssim 10^{10}$ \msun; 
disc-dominated ($0.1<\bt\leq0.45$) at intermediate masses, $10^{10}\lesssim \ms/\msun\lesssim 8\times 10^{10}$;
and bulge-dominated  ($\bt>0.45$) at large masses, $\ms\gtrsim 8\times 10^{10}$ \msun. 
In the $1<z<3$ period, galaxies with $\bt\leq 0.45$ dominate by far at masses below $\ms\sim 10^{11}$ \msun, 
and for $z>2$, galaxies with $\bt\leq 0.1$ dominate at all masses.

\item For massive galaxies, $\ms>10^{11}$ \msun, the fraction of bulge-dominated systems rises 
systematically with cosmic time, becoming the dominant population at lower redshifts. 
The setting of the bulge-dominated (early-type) population follows on average a morphology and 
a mass assembly downsizing trend: the more massive the present-day early-type galaxy, the
earlier it suffered the morphological transformation and the earlier it assembled most of its mass.
Contrary to semi-analytical model predictions, this is in agreement with observations. 
The dominant channel of bulge growth for the massive galaxies is the acquisition
of stellar mass from the secondary(ies) in major mergers.

\item At $z>1$, the galaxies that become bulge-dominated (their \bt\ ratios overcome 0.45) 
are on average still actively growing in mass, presumably by in situ star formation. At $z\lesssim0.5$,
a significant fraction of the galaxies in transition to bulge dominion are already passive, in such
a way that the merger(s) they subsequently suffer to become bulge-dominated are presumably dry. 

\item  The predicted local bulge demographics as a function of mass is in agreement with 
observations (see also Paper I). At higher redshifts, 
the few observational studies available use 
instead of the \bt\ ratio, other morphological definitions. 
Taking this into consideration, we have compared our models to 
observations from \citet{Oesch+2010}, \citet{Kovac+2010}, 
\citet[][these authors actually measure the \bt\ ratio]{Bruce+2012}, 
\citet{Buitrago+2013}, and \citet{Mortlock+2013}. We found that, 
within the large observational systematic and statistical uncertainties,
the trends of the fractions of galaxies with a given \bt\ as
a function of (\ms,$z$) are in agreement.
It is particularly remarkable the excellent agreement with \citet{Buitrago+2013} in 
the fractions of bulge- and disc-dominated massive galaxies from $z=3$ to $z\sim0$.

\item According to our merger-driven bulge growth predictions, the \lcdm\ scenario does not face
the problem of producing a deficit of bulgeless or disc-dominated galaxies
at intermediate/low masses. If any, it seems to predict slightly more of such galaxies with respect to local and
high $z$ observations. Thus, there is room for including intrinsic (not induced by mergers) 
mechanisms of bulge growth, for instance, the central migration of gas-rich clumps produced by 
instabilities of the gaseous discs at $z\sim 1-2$, and the secular bar formation and dissolution in 
evolved stellar discs.    

We note that our semi-empirical model results depend 
on the way the SHMR evolves. For a ``slowly evolving'' SHMR \citep[e.g.][]{Behroozi+2013-L}, 
the predicted bulge demographics and evolution of the morphological mix are in tension with 
observations:  the predicted \bt\ ratios of low mass galaxies are too high, and the bulge assembly 
of bulges in massive galaxies is predicted to occur too early.

\end{itemize}

\section*{Acknowledgments}

We are grateful to the Referee for useful comments that have helped to improve the presentation of 
the paper. VA acknowledges PAPIIT-UNAM grant IA-100212 and CONACyT grant 167332 (Ciencia B\'asica)
for partial support.
JZ is supported by the University of Waterloo and the Perimeter Institute for Theoretical Physics. Research 
at Perimeter Institute is supported by the Government of Canada through Industry Canada and by the Province 
of Ontario through the Ministry of Research and Innovation. The Dark Cosmology Centre is funded by the 
DNRF. IL is supported by a DGAPA-UNAM Postdoctoral Fellowship.

\bigskip

\appendix

\section{The semi-empirical relations}
 
The shape of the SHMR used here has been proposed by \citet{Behroozi+2010} on the basis of 
its ability to map the halo mass function into a Schechter-like Galaxy Stellar Mass Function down 
to $\ms\sim 10^9$ \msun. \citet{Behroozi+2010} constrained with observations the parameters of this SHMR
from $z=0$ to $z=1$ and from $z=1$ to $z=4$ independently. 
\citet{Firmani+2010} slightly modified the \citet{Behroozi+2010} SHMR in order to describe
its evolution in a continuous way from $z=0$ to $z=4$. The analytical formula is as follows:
\begin{displaymath}
 \log(\mh(\ms)) = \hspace{0.65\columnwidth}
 \end{displaymath}
 \vspace{-3ex}
\begin{eqnarray}
\quad \log(M_1) + \beta\,\log\left(\frac{\ms}{\ms_{,0}}\right) +
 \frac{\left(\frac{\ms}{\ms_{,0}}\right)^\delta}
{1 + \left(\frac{\ms{,0}}{\ms}\right)^{\gamma}} - \frac{1}{2}.
\label{Mh_Ms}
\end{eqnarray}
The dependence on $z$ is introduced in the parameters of eq. (\ref{Mh_Ms}) as:
\begin{eqnarray}\label{zevolution}
\log(M_1(a)) & = & M_{1,0} + M_{1,a} \, (a-1), \nonumber\\
\log(\ms_{,0}(a)) & = & \ms_{,0,0} + \ms_{,0,a} \, (a-1)+ \chi \left( z \right), \nonumber \\
\beta(a) & = & \beta_0 + \beta_a \, (a-1),\\ 
\delta(a) & = & \delta_0 + \delta_a \, (a-1), \nonumber \\
\gamma(a) & = & \gamma_0 + \gamma_a \, (a-1),  \nonumber
\end{eqnarray} 
where $a=1/(1+z)$ is the scale factor.
The function $\chi \left( z \right)$ controls the change with $z$ of the peak value of the \ms-to-\mh\ ratio. 
\citet{Firmani+2010} defined $\chi \left( z \right)$ in order to roughly reproduce the peak evolution 
found in \citet{Behroozi+2010}:
\begin{eqnarray}
\chi \left( z \right) = -\chi_{0} \ z (1-0.378 z (1-0.085 z));
\end{eqnarray}
if $\chi =0$, then the peak of the \ms-to-\mh\ ratio remains the same at any $z$.
The first two formulae in eq. (\ref{zevolution}) control the position of the \ms/\mh\ peak at each $z$, 
while the last three control the shape of the \ms/\mh--\mh\ 
curves. The set of parameter values reported in \citet{Firmani+2010} and used here are reproduced in the
second column of Table \ref{parameters}. 

In Section \ref{slowSHMR} we experimented with a slowly evolving SHMR, based on the recent constraints
by \citet{Behroozi+2013-L}. By keeping the same parametrization given by eqs. (\ref{Mh_Ms}) and (\ref{zevolution}),
the parameter values that closely reproduce the \citet{Behroozi+2013-L} SHMRs are reported
in the third column of Table \ref{parameters}.

\begin{table}
\begin{center}
\caption{Parameters for the moderately 
and slowly evolving SHMRs}
\label{parameters}
\begin{tabular}{lrr}
\hline
\hline
Parameter & Moderate & Slow \\
\hline
$\ms_{,0,0}$  &  $10.70$  & 10.73 \\
$\ms_{,0,a}$  &  $-0.80$   & $-0.30$ \\
$M_{1,0}$      &  $12.35$  & 12.45 \\
$M_{1,a}$      &  $-0.80$   & $-0.44$ \\
$\beta_0$      &  $0.44$    & 0.46 \\
$\beta_a$      &  $0.00$    & 0.00 \\
$\delta_0$     &  $0.48$    & 0.48 \\
$\delta_a$     &  $-0.15$   & $-0.95$ \\
$\gamma_0$  &  $1.56$   & 0.96 \\
$\gamma_a$  &  $0.00$   & $-0.70$ \\
$\chi_{0}$        &  $0.181$    & $0.018$ \\
\hline
\end{tabular}
\end{center}

\end{table}


The \ms--\mg\ relation and its change with redshift used in our
model (see Section 2.2) has been proposed by \citet{Stewart-09} 
as a fit to the available data at $z\sim 0$ and at higher redshifts: 
\begin{eqnarray}
\frac{\mg}{\ms}(z) = 0.04 \left(\frac{\ms}{4.5\times 10^{11}\msun}\right)^{-\alpha(z)}, 
\end{eqnarray}
where $\alpha(z)= 0.59(1+z)^{0.45}$. For small masses, \mg/\ms\ can be very large, particularly at higher
redshifts. Given the high degree of observational uncertainty in this regime, 
we opt to set $\mg/\ms\le 100$, which is the maximum observed value  
reported in \citet{Stewart-09}.

\bibliography{./lit-2}

\label{lastpage}

\end{document}